\documentclass[aps,12pt,nofootinbib,showpacs,showkeys,preprintnumbers,amsmath,amssymb]{revtex4-1}		

\usepackage{amsmath,amsfonts,amssymb,amsthm,bm,bbm,bbold,braket,color,dsfont,fancybox,hyperref,
srcltx,subcaption,graphicx,ucs,yfonts,cancel,xfrac,verbatim,xcolor,slashed}
        
\usepackage{float}

\usepackage[section]{placeins}

\usepackage{lineno}

\begin{document}

\title{Fermion Dark Matter in the Vector Scotogenic Model: A Survey of Signatures}

\author{Paulo Areyuna C.$^{1}$}
\email{paulo.areyuna@sansano.usm.cl}

\author{Jilberto Zamora-Saa$^{1,2}$}
\email{jilberto.zamora@unab.cl}

\author{Alfonso R. Zerwekh$^{1,3,4}$}
\email{alfonso.zerwekh@usm.cl}

\affiliation{$^1$Millennium Institute for Subatomic physics at high
energy frontier - SAPHIR, Fernandez Concha 700, Santiago, Chile.}
\affiliation{$^2$Center for Theoretical and Experimental Particle Physics - CTEPP, Facultad de Ciencias Exactas, Universidad Andres Bello, Fernandez Concha 700, Santiago, Chile.}
\affiliation{$^3$ Departamento de Física, Universidad Técnica Federico Santa María Casilla 110-V, Valparaíso, Chile.}
\affiliation{$^4$Centro Científico - Tecnológico de Valparaíso, Casilla 110-V, Valparaíso, Chile}

\begin{abstract}
In this work, we have studied the Vector Scotogenic Model in the context of the Dark Matter problem. Due to unitarity considerations, we have focused on the scenario with fermion dark matter, finding out that co-annihilations play a fundamental role in achieving dark matter relic abundance. Moreover, the coannihilation effects allow to separate the parameter space into two regions with different phenomenology. In addition,
we have studied the detection prospects of these regions separately, focusing on signatures that can appear in lepton flavor violating decays, indirect and direct searches, and the production of these new particles at collider facilities.
\end{abstract}
\maketitle

\section{Introduction}
The standard model of particle physics (SM) has been highly successful in explaining fundamental interactions, but it has limitations in accounting for certain phenomena, such as Dark Matter (DM) and neutrino mass generation. Although the amount of DM in the Universe is well known \cite{PLANCK}, its nature remains as a mystery. On the other side, there is no evidence on the existence of right handed neutrinos, making difficult to explain neutrino masses by electroweak symmetry breaking. 
One intriguing possibility is to connect the apparently independent problems of DM and neutrino mass generation. The first attempt to solve these two problems in the same framework was the scotogenic model Ref. \cite{Ma:2006km}, which is an extension to the SM by a singlet fermion and a massive scalar $SU(2)_L$ doublet. Under this framework, the neutrinos acquire mass via radiative processes involving the doublet components. Across the years, many variations of the model have been studied. Among these model variants, we focus our attention in the Vector Scotogenic Model \cite{masses_and_mixings,dong2021}. In this variant of the scotogenic paradigm, the doublet has spin 1. This change implies that the singlet fermion must be left-handed. We have studied this model in the past, in the context of collider probes of new physics \cite{colliderLHNL}. In this work, we develop a comprehensive analysis based on dark matter phenomenology, considering its production in the early universe and the detection prospects nowadays.
This paper is structured as follows: In Section \ref{sec:model} we review the main features of the model. In Section \ref{sec:early_universe} we present a preliminary scan varying only key parameters in the model, in order to understand the production mechanism in the early universe, and the corresponding implications. In Section \ref{sec:relic}, we generalize the results from Section \ref{sec:early_universe} for the relic abundance. In Section \ref{sec:lfv}, we show the constraints arising from lepton flavor violating processes. In Sections \ref{sec:indirect} and \ref{sec:direct} we discuss the discovery potential of these new particles by means of astrophysical signatues: such as indirect detection of dark matter annihilation and anomalous interactions with nuclei for direct detection. The study of astrophysical signatues is complemented with the discovery prospects at collider facilities, detailed in Sections \ref{sec:vector_production} and \ref{sec:collider}. Finally, we summarize our conclusions in \ref{sec:conclusions}

\section{The model}\label{sec:model}
The Vector Scotogenic Model is an extension to the SM composed by a massive vector doublet, defined as:
\begin{equation}
    V_{\mu}=\begin{pmatrix}V_\mu^+\\\frac{1}{\sqrt{2}}(V_\mu^1+iV_\mu^2)\end{pmatrix}\sim (1,2,1/2),
\end{equation}
 and a left-handed singlet fermion $N_{L}\sim (1,1,0)$, which is assumed to be a Majorana particle. The SM group is extended by a $Z_2$ symmetry in which the new particles are odd and all the SM particles are even, assuring stability of the dark sector. The vector doublet presents electroweak interactions described by the following lagrangian:
\begin{equation}
\begin{split}
    \mathcal{L}_{V}=&-\frac{1}{2}(D_\mu V_\nu -D_\nu V_\mu)^\dagger (D^\mu V^\nu- D^\nu V^\mu)+M_V^2 V_\mu^\dagger V^\mu-\frac{1}{\xi}(D_\mu V^\mu)^\dagger(D_\nu V^\nu)
    \\&+\kappa\big[i\frac{g'}{2}V_\mu^\dagger B^{\mu\nu}V_\nu+ig V_\mu^\dagger W^{\mu\nu}V_\nu\big] -\alpha_2(V_\mu^\dagger V^\mu)(V_\nu^\dagger V^\nu)-\alpha_3(V_\mu^\dagger V^\nu)(V_\nu^\dagger V^\mu)\\&-\lambda_2(\Phi^\dagger \Phi)(V_\mu^\dagger V^\mu)-\lambda_3(\Phi^\dagger V_\mu)(V^{\mu\dagger}\Phi)-\frac{\lambda_4}{2}[(\Phi^\dagger V_\mu)(\Phi^\dagger V^\mu)+(V^{\mu\dagger}\Phi)(V_\mu^\dagger \Phi)],
    \end{split}
\end{equation}
where $D_\mu$ stands for the covariant derivative, $W^{\mu\nu}$ and $B^{\mu\nu}$ are the field strengths of $SU(2)_L$ and $U(1)_Y$, respectively, and $\Phi$ is the SM Higgs doublet.One interesting feature about this lagrangian is the presence of non minimal gauge interactions, described by the parameters $1/\xi$ and $\kappa$. These terms should play a relevant role for  UV completions of the model. Also, the couplings $\alpha_2,\alpha_3$ describe pure interactions among the vector fields. These
 self-interacting terms are not relevant for the phenomenological aspects described in this paper, therefore from now on we will not consider them, However,
self-interacting particle dark matter can be relevant in related fields such as astrophysical
structures \cite{Tulin:2017ara}.

In addition,the new fermion interactions are described by the following lagrangian:

\begin{equation}
    \mathcal{L}_{HNL}=\frac{1}{2}(i\bar{N}_L^c\gamma^\mu \partial_\mu N_L -M_N\bar{N}_L^cN_L)-\sum_{k=\{e,\mu,\tau\}}\beta_k \bar{L}_k \gamma^\mu\tilde{V_\mu} N_L  +\text{h.c.},
\end{equation}
with $\tilde{V_\mu}=i\sigma_2 V_\mu^*$. 
The mass spectrum of the theory can be found in Table \ref{masses}.
%%%% deberir ir en la intro?
This model was firstly studied in Ref. \cite{masses_and_mixings} as an extension to the Vector Doublet Dark Matter Model (VDDMM) studied in Ref. \cite{vector_dm}, finding out that the model can account for neutrino masses. After that, the model was probed in the context of the muon $g-2$ anomaly by the authors of Ref. \cite{dong2021}. The capability of the model to solve several problems in theoretical physics motivated us to study the model in the context of collider probes for new physics (Ref. \cite{colliderLHNL}). In the present work, we focus on dark matter phenomenology arising from this model.

%%%
which presents two dark matter candidates: the singlet fermion and the neutral component of the vector doublet. However, as stated in Ref. \cite{vector_dm}, perturbative unitarity is achieved when $M_{V^+}\approx M_{V^1}\approx M_{V^2}$. In this kinematical regime, the cutoff scale of the theory spans between $3$ to $10$[TeV], assuring consistency for vector masses below the TeV scale. Besides, results in neutrino mass generation \cite{dong2021}  show that the mass difference between $V^1$ and $V^2$ should be small in order to get  $\beta$ couplings around $\sqrt{4\pi}$.
Since $M_N$ is not constrained by perturbative unitarity, we will consider the simplified case of $M_{V^+}= M_{V^1}=M_{V^2}$ and $M_N\leq M_{V^+}$, making $N_L$ the only dark matter candidate. 
 It's worth mentioning that these fermions don't carry lepton number, therefore it's possible to add an arbitrary number of new singlet fermions without introducing anomalies. Indeed, at least two singlet fermions are necessary to reproduce the neutrino mixing matrix. Since we are considering the physics case of fermion DM, we will assume the existence of additional fermions which are much heavier than the vector fields, and are responsible for neutrino mass generation. In such a scenario, the coupling of the lightest fermion to leptons is independent of the PMNS matrix, opening the phenomenology and allowing to scan over different values of these couplings. Note that, in principle, the $\beta$ couplings could be complex, inducing CP violating phases. Since neutrino mass generation is not possible in our simplified setup with 1 fermion, we restricted ourselves to the scenario where these couplings are real. However, we recognize the importance of CP phases for future studies where the PMNS matrix is taken into account. In light of these assumptions, we will consider two benchmark cases, $M_{V^+}=200$[GeV] (the low mass regime from now on) and $M_{V^+}=800$ (the high mass regime from now on). The low mass regime is motivated by collider limits while the upper bound is the maximum value consistent with perturbative unitarity. Finally, we define the following control values for the parameters involving the vector doublet interactions:
\begin{equation}
    \frac{1}{\xi}=0,\quad \kappa=-1,\quad \lambda_L\equiv \lambda_2+\lambda_3+\lambda_4=5.
\end{equation}
 These choices are motivated by the following reasons: the parameters $1/\xi$ and $\kappa$ introduce modifications on the gauge interactions of the new vector fields, the choice presented here makes $V^+$ more similar to the $W$, allowing the use of previous results on radiative processes relating electroweak observables (as will become evident on Section \ref{sec:lfv}). On the other hand, a positive value of $\lambda_L$ helps to push the cutoff scale to higher values, allowing to perform phenomenological calculations safely.

\begin{table}[!h]
    \centering
    \begin{tabular}{|c|c|}
    \hline
    particle & physical mass \\
    \hline
    $V^+$     & $M_{V^+}\equiv\sqrt{\frac{1}{2}(2M_V^2-v^2\lambda_2)}$\\
    $V^1$    &$M_{V^1}\equiv\sqrt{\frac{1}{2}(2M_V^2-v^2[\lambda_2+\lambda_3+\lambda_4])}$\\
    $V^2$    & $M_{V^2}\equiv\sqrt{\frac{1}{2}(2M_V^2-v^2[\lambda_2+\lambda_3-\lambda_4])}$\\
    $N_L$     &   $M_N$\\
    \hline
    \end{tabular}
    \caption{Mass spectrum of the Vector Scotogenic Model after electroweak Symmetry Breaking}
    \label{masses}
\end{table}

\section{Early universe dynamics}\label{sec:early_universe}
\begin{figure}[!h]
    \centering
    \includegraphics[width=0.8\textwidth
]{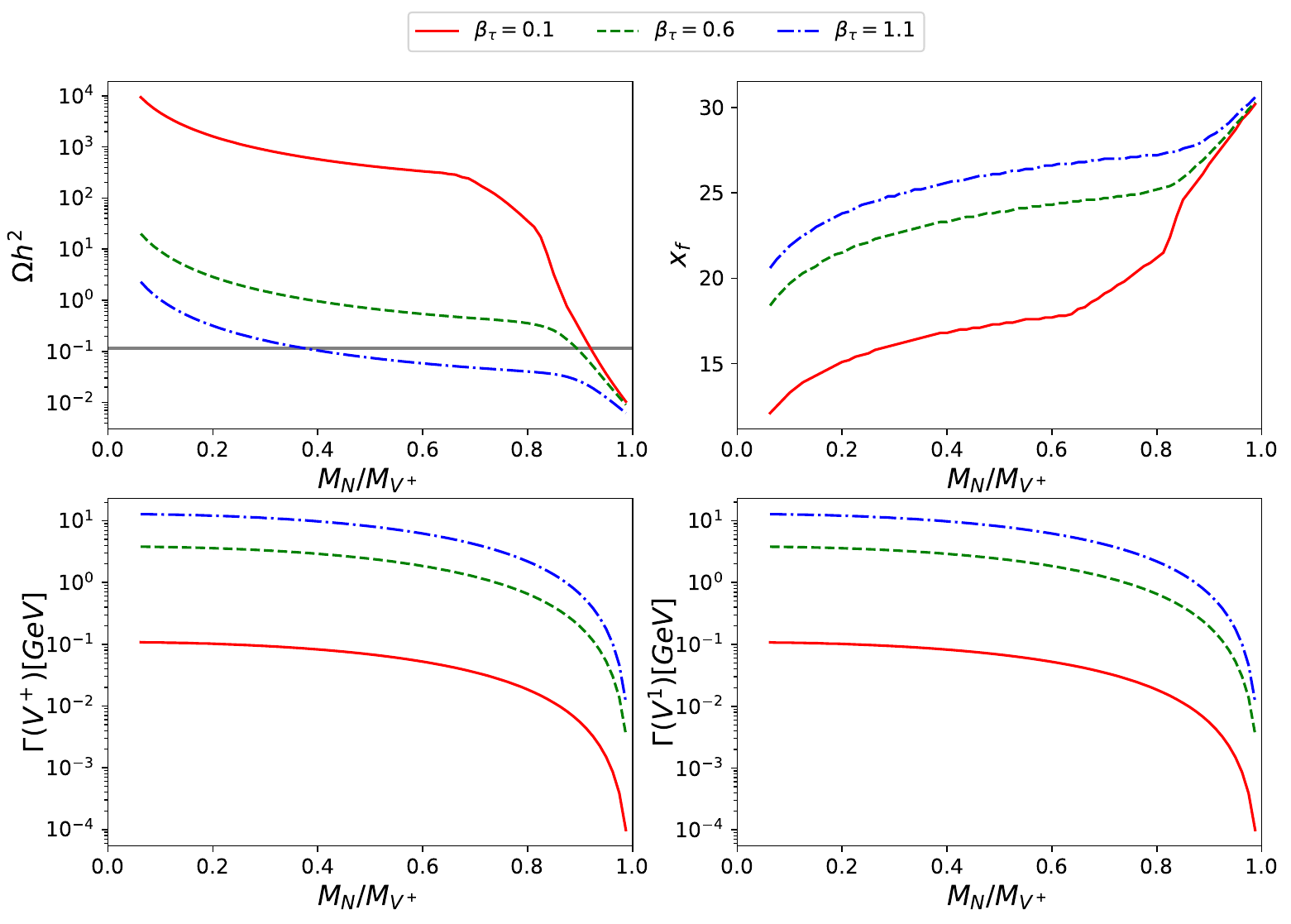}
    \caption{Relevant observables related to the early universe dynamics as a function of $M_N/M_V^{+}$ and different values of $\beta_\tau$. For these plots, we considered $M_{V^+}=800$[GeV] and $\beta_e=\beta_\mu=0$. The solid horizontal gray line in the first panel shows the measured value by PLANCK.}
    \label{ratio_bench}
\end{figure}

The thermal equilibrium of the dark sector in the early universe is determined by three types of processes: direct annihilation of dark fermion pairs, coannihilation involving one fermion and the new vector field, and the annihilation of the vector field. The relative contribution of each channel depends on the kinematical regime and the choice of the $\beta$ couplings. In order to simplify the analysis, we considered firstly the special case when $\beta_e=\beta_\mu=0$. We have used micrOMEGAs \cite{micromegas1,micromegas3} to carry out a scan over different values of $M_N, M_{V^+}$ and $\beta_\tau$. As can be seen in Figure \ref{ratio_bench},  dark matter relic density presents a strong dependence on $\beta_\tau$. When this quantity is small, the early universe dynamics is dominated by coannihilation and pure vector annihilation, however  the contribution of these processes to the thermally averaged annihilation cross section is suppressed by the Boltzmann factor $B\approx e^{-\frac{2(M_{V^+}-M_N)}{T}}$, becoming relevant only in the nearly degenerate regime $M_N\sim M_{V^+}$, and allowing to saturate the relic abundance even for small $\beta$ values. This behavior is analogous to the one reported in the scalar scotogenic models  \cite{scalar_scoto,Baumholzer:2019twf}. It's worth mentioning that in the regime where $M_N\sim M_{V^+}$, the relic density is achieved by the vector decay, under a freeze-in mechanism where the source of dark matter is in thermal equilibrium (which would correspond to a late freeze-out according to Ref. \cite{freezeinfreezeout}). Finally, we notice that the production mechanism of dark matter can affect the abundance of SM leptons, however, the description of these effects come from a detailed study of the Boltzmann equations for dark matter and neutrinos, which are coupled. A detailed description of these effects is beyond the scope of this work and should be studied elsewhere.

\section{Relic density}\label{sec:relic}
\begin{figure}[!h]

    \begin{subfigure}{0.495\textwidth}
    \includegraphics[width=\textwidth]{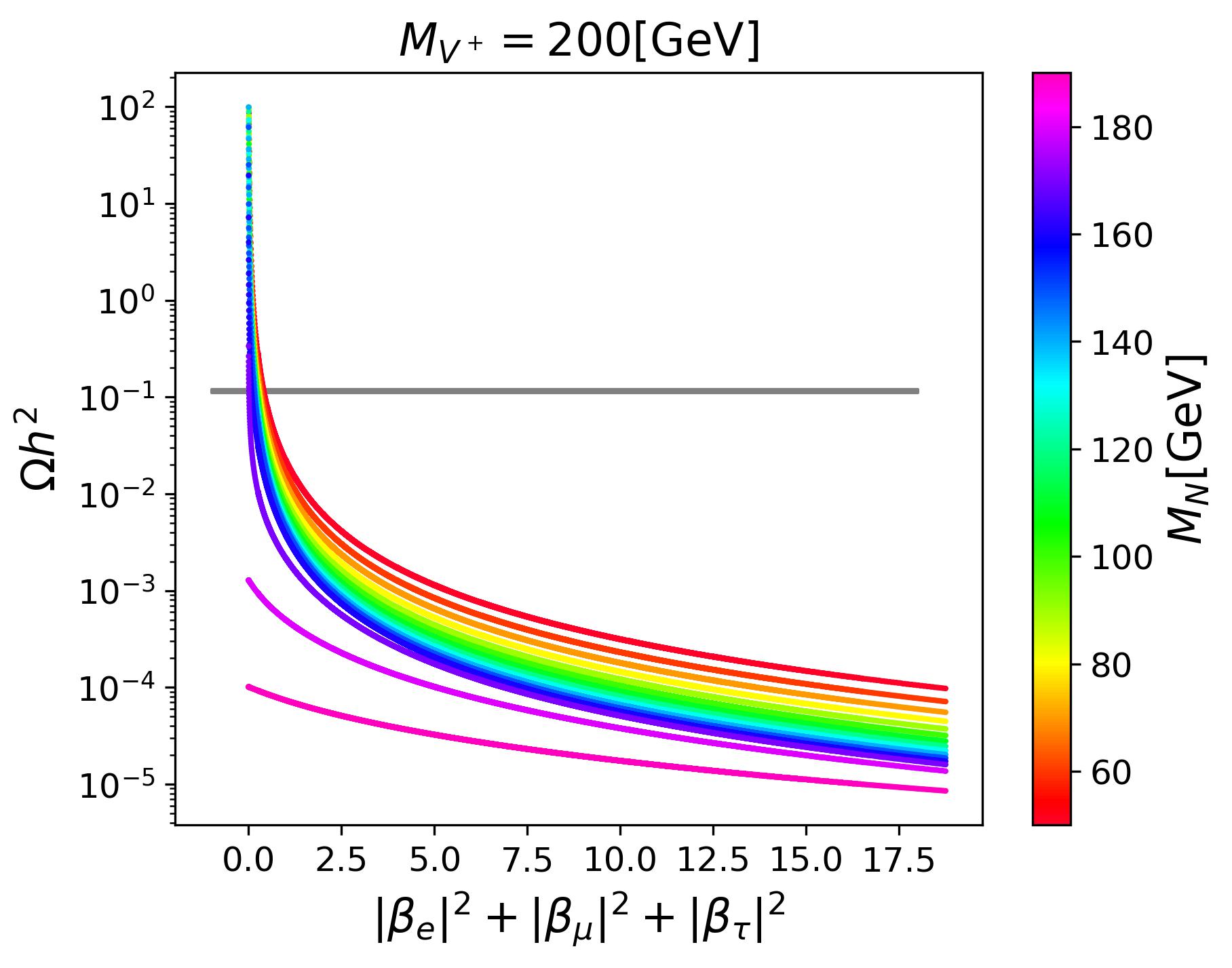}
    \caption{}
\end{subfigure}
    \begin{subfigure}{0.495\textwidth}
    \includegraphics[width=\textwidth]{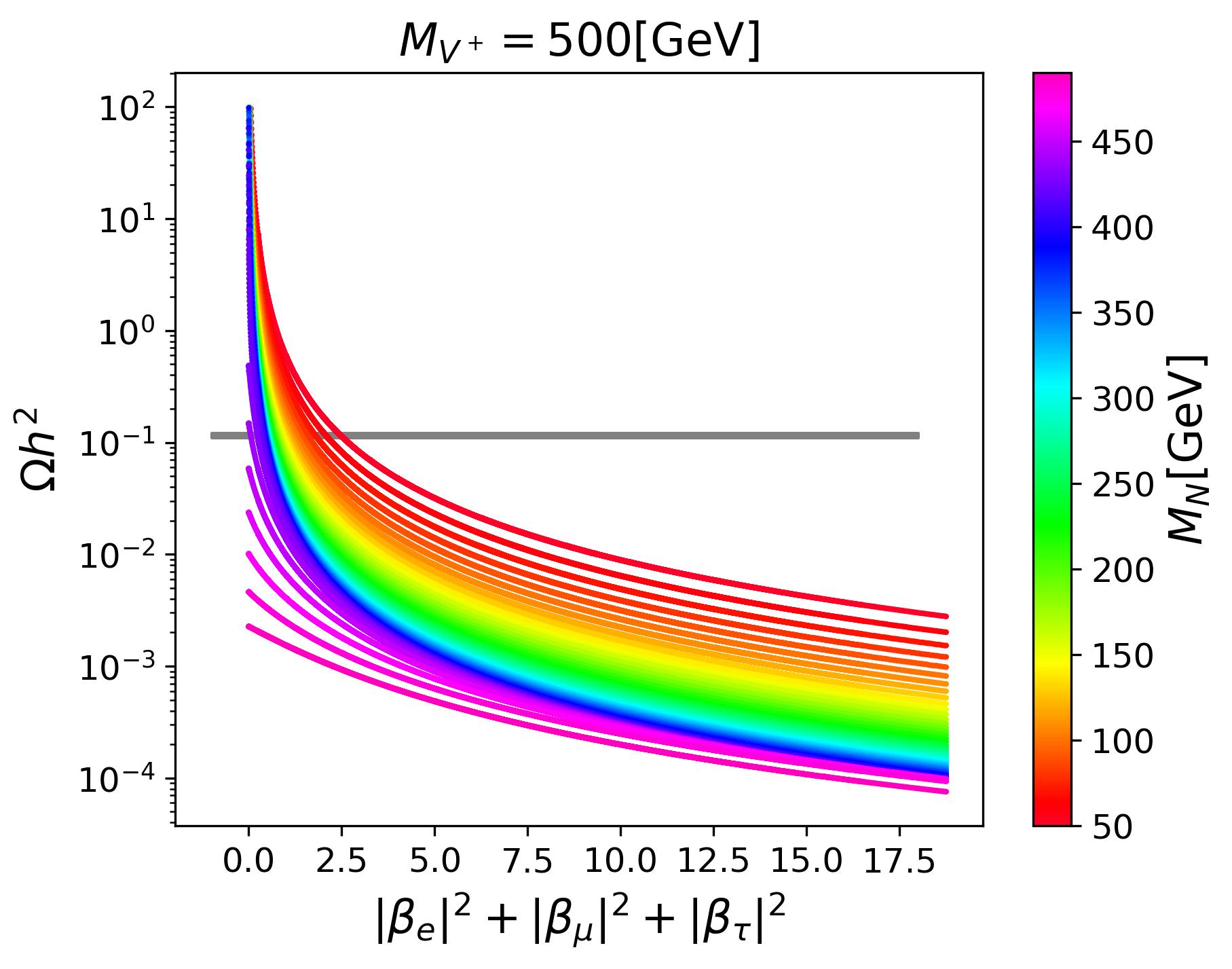}
    \caption{}
\end{subfigure}
    \begin{subfigure}{0.495\textwidth}
    \includegraphics[width=\textwidth]{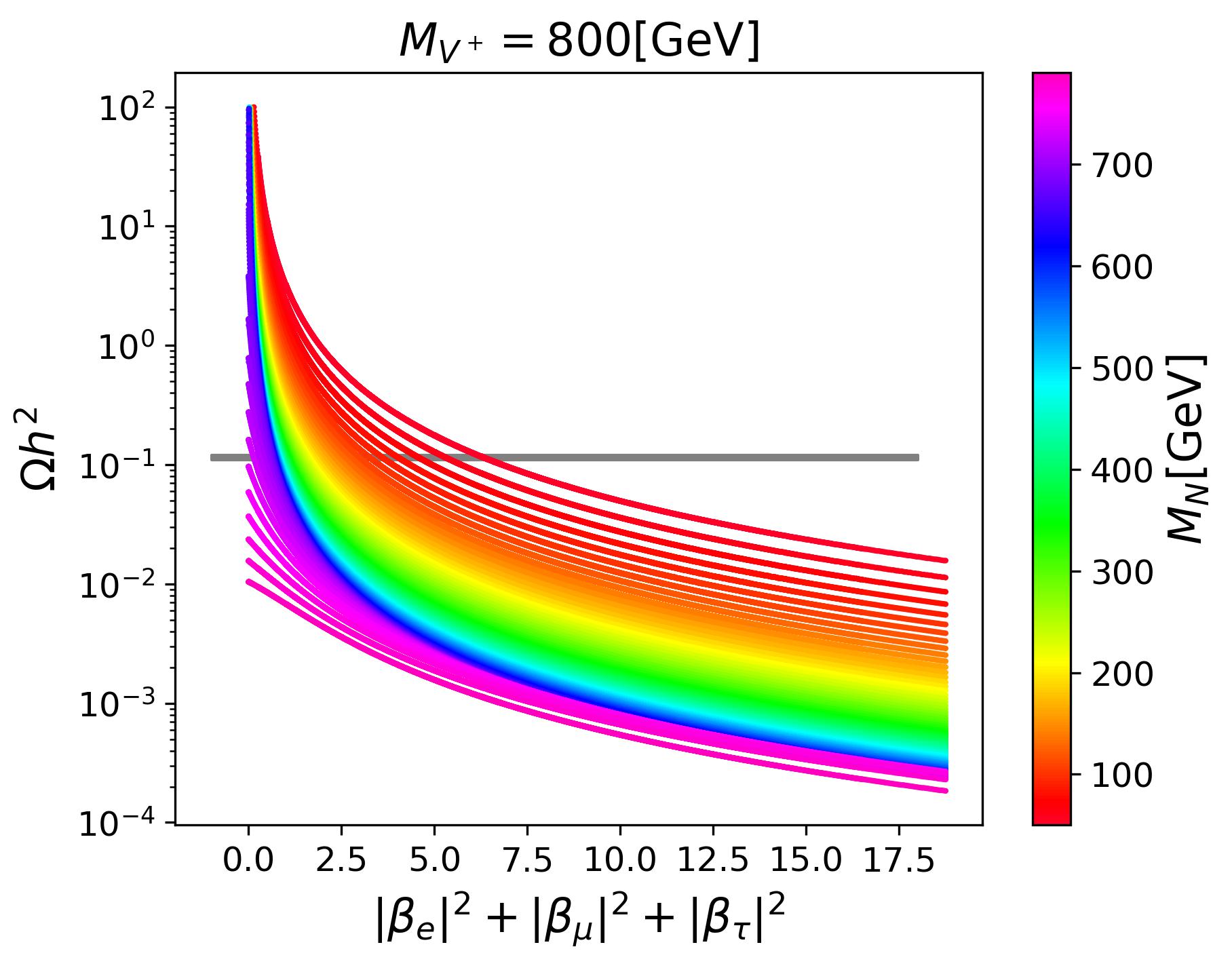}
    \caption{}
\end{subfigure}

\caption{Relic density dependence on the squared sum of the $\beta$ couplings for different values of $M_N$ and $M_{V^+}$.}
\label{relic}

\end{figure}

\begin{figure}[!h]
    \begin{subfigure}{0.6\textwidth}
            \includegraphics[width=\textwidth]{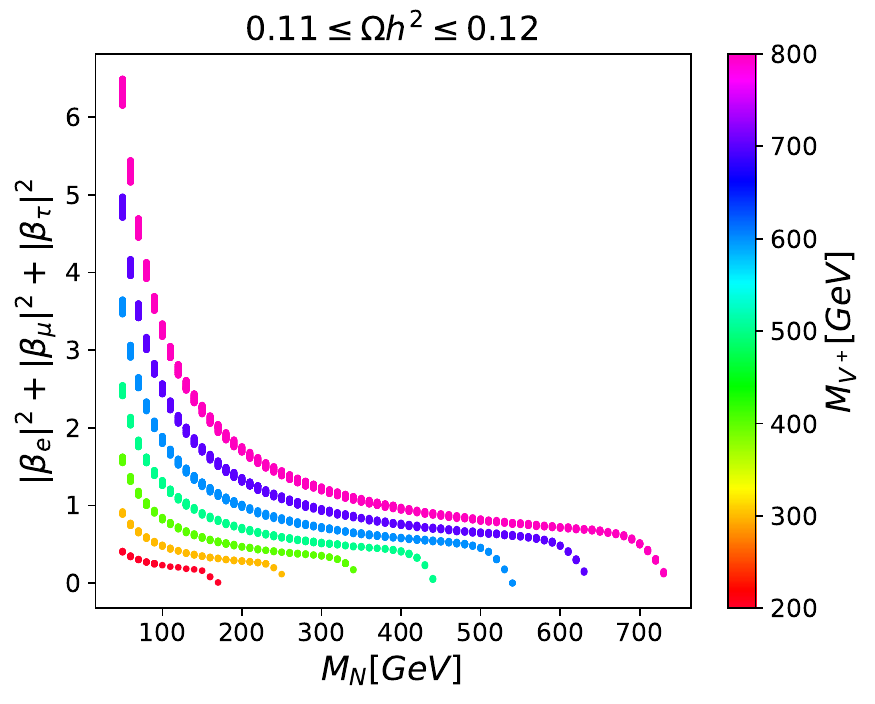}
            \caption{}
    \end{subfigure}
    \begin{subfigure}{0.6\textwidth}
            \includegraphics[width=\textwidth]{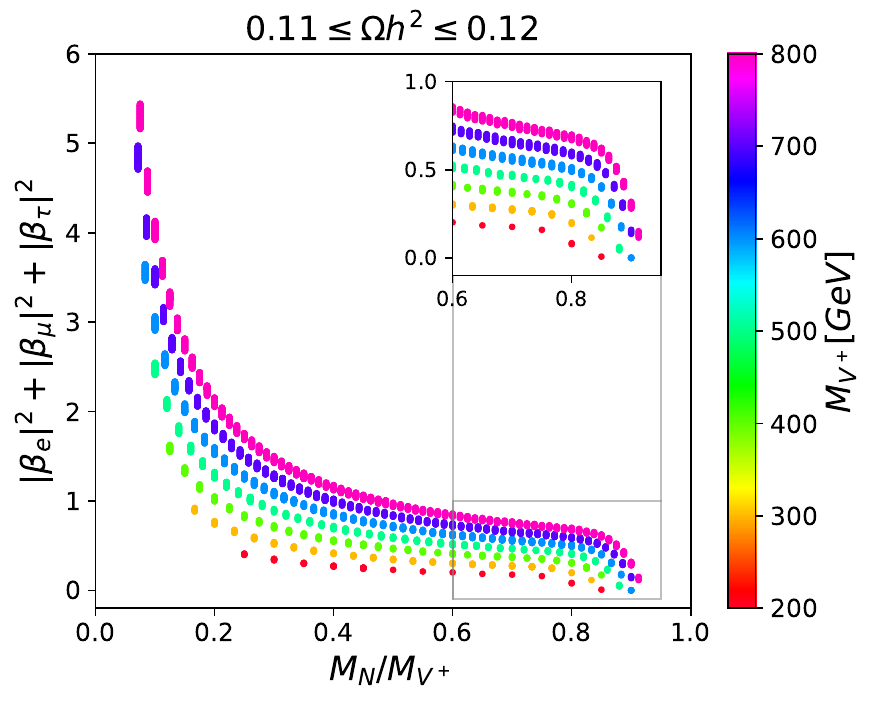}
            \caption{}
    \end{subfigure}

    \caption{Parameter space points that saturate the dark matter relic abundance.}
    \label{relic_sat}
\end{figure}

In this section we generalize the results from the previous section considering different values for $\beta_e$ and $\beta_\mu$. We have performed a scan over the parameter space and computed the DM relic density.  We constrained the parameter space considering the PLANCK measurement \cite{PLANCK} which is reported as $\Omega h^2=0.12\pm 0.001$. As can be seen in Figure \ref{relic}, the relic density is inversely proportional to the square sum of the couplings, presenting a strong suppression. This behavior can be used to define lower limits on the couplings for a given choice of $M_N$ and $M_{V^+}$. However, this is not possible when $ M_{V^+}\geq 0.8 M_N$, because in this case the thermal relic density is dominated by coannihilation processes involving the vector states, and therefore the $\beta$ couplings don't play a relevant role. We made a focus on the region of the parameter space satisfying $0.11\leq\Omega h^2\leq 0.12$, where the model can saturate relic density up to $\sim 90\%$ of the total value. This region can be seen in Figure \ref{relic_sat}, and it's easier to note the change in channel contribution for $M_{V^+}\geq 0.8 M_N$. It's worth mentioning how the saturation region differs from the approximated results in our previous work \cite{colliderLHNL}, a detailed discussion about this discrepancy can be found in \ref{sec:ap_2}.

\section{Lepton Flavor Violation}\label{sec:lfv}
\begin{figure}[!h]

    \begin{subfigure}{0.495\textwidth}
    \includegraphics[width=\textwidth]{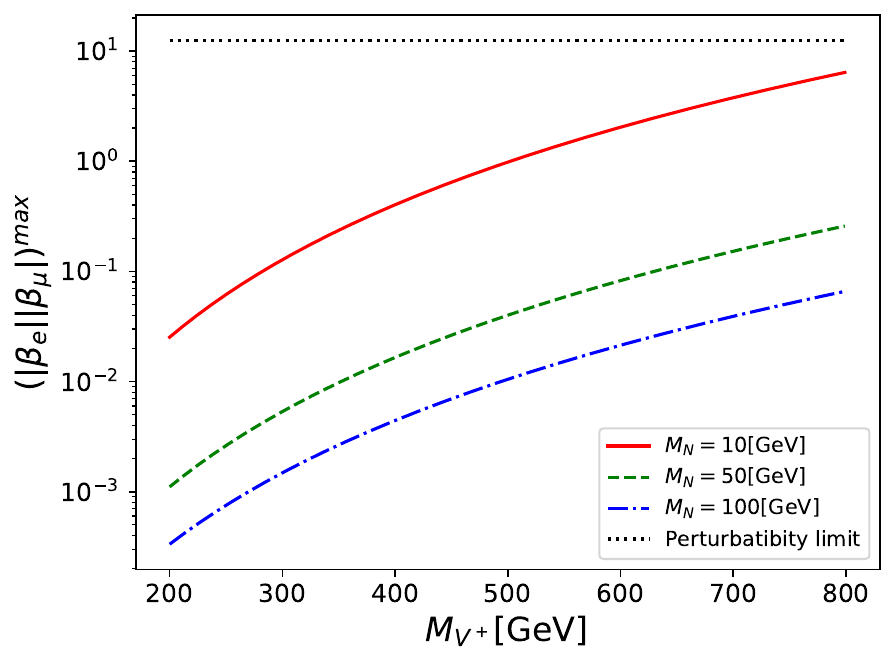}
    \caption{}
\end{subfigure}
    \begin{subfigure}{0.495\textwidth}
    \includegraphics[width=\textwidth]{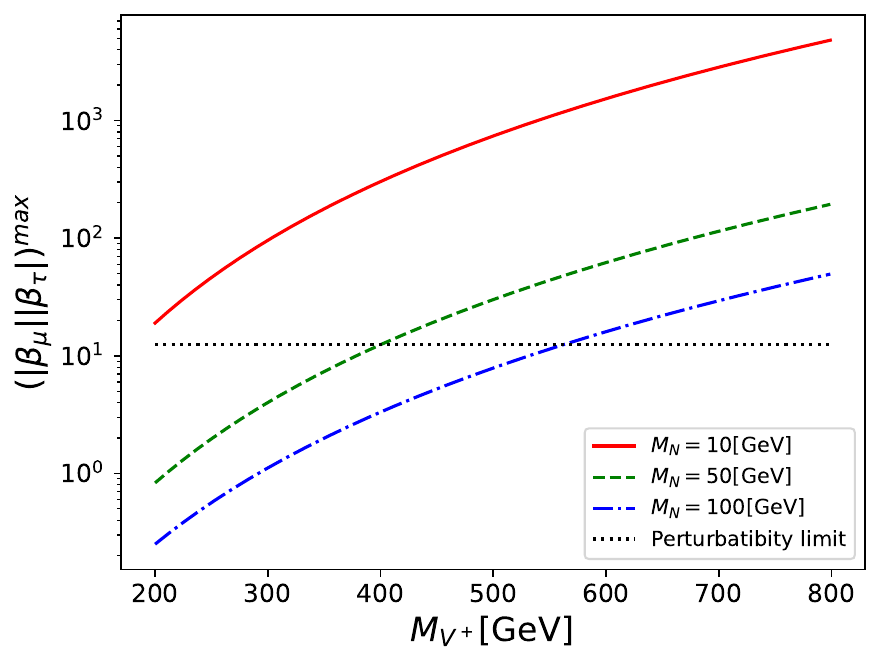}
    \caption{}
\end{subfigure}
    \begin{subfigure}{0.495\textwidth}
    \includegraphics[width=\textwidth]{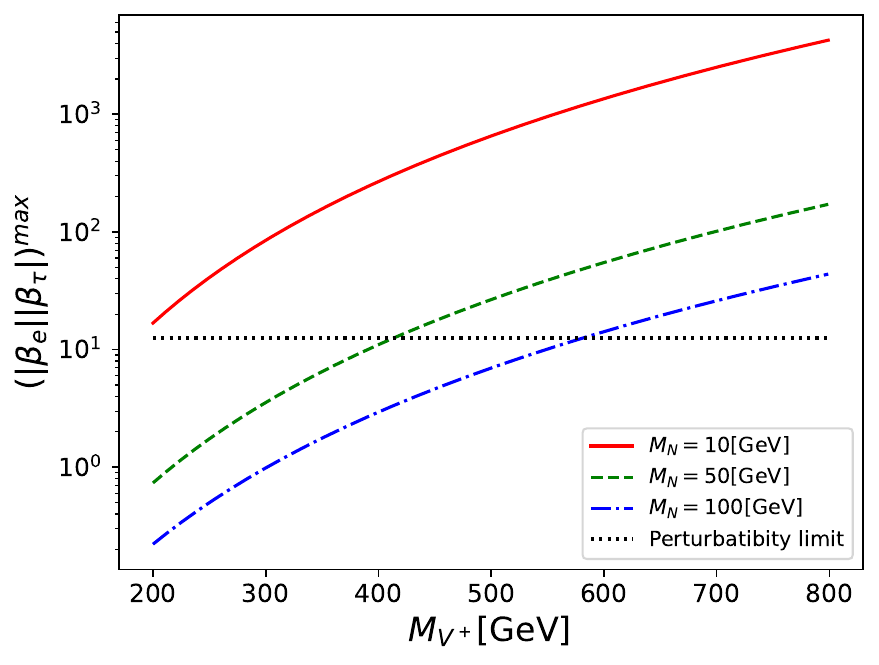}
    \caption{}
\end{subfigure}

    \caption{Upper limits for the product $|\beta_i||\beta_j|$ ($i,j=e,\mu,\tau$) for some reference values of $M_N$ and $M_{V^+}$. We notice that limits are relaxed for higher values of $M_{V^+}$. }
    \label{lfv}

\end{figure}

While the DM relic abundance depends on the squared sum of the couplings, these are constrained by  lepton flavor violation (LFV) decays. This type of process is very rare and the upper bound on these quantities can be seen in Table \ref{lfv_pdg}. According to Ref. \cite{ilakovac} the branching fraction for charged LFV decay has the following form:
\begin{equation}
    \text{Br}(l_i\to l_j\gamma)_{\kappa=-1}=\frac{|\beta_i|^2|\beta_j|^2g_w^2s_w^2m_i^5}{64(4\pi)^4M_{V^+}^4\Gamma_i}|G\left(M_N^2/M_{V^+}^2\right)|^2,
\end{equation}
with 
\begin{equation}
    G(x)=-\frac{2x^3+5x^2-x}{4(1-x)^3}-\frac{3x^3}{2(1-x)^4}\ln x.
\end{equation}
This expression can be used to define limits on the product of couplings, for instance limits over $|\beta_e||\beta_{\mu}|$ as is shown in Figure \ref{lfv}. It's worth mentioning that these limits are valid only when $\kappa=-1$, a different value for this parameter should affect the $G(x)$ function. Under this parameter setting, the $V^+$ magnetic moment has the same form as the $W^+$ magnetic moment, allowing us to use the results from Ref. \cite{ilakovac}. This similarity could be relevant for UV completions of the model considering a larger gauge group. 

\begin{table}[!h]
    \centering
    \begin{tabular}{|c|c|}
    \hline
     process    &  branching fraction upper limit \\
     \hline
       $\mu \to e \gamma$  & $< 4.2\times 10^{-13}$\\
    $\tau \to e \gamma$  & $< 3.3\times 10^{-8}$\\
    $\tau \to \mu \gamma$  & $< 4.2\times 10^{-8}$\\
    \hline
    \end{tabular}
    \caption{Current limits on charged LFV decays, taken from Ref. \cite{pdg2022}.}
    \label{lfv_pdg}
\end{table}

\begin{figure}[!h]
    \begin{subfigure}{0.495\textwidth}
            \includegraphics[width=\textwidth]{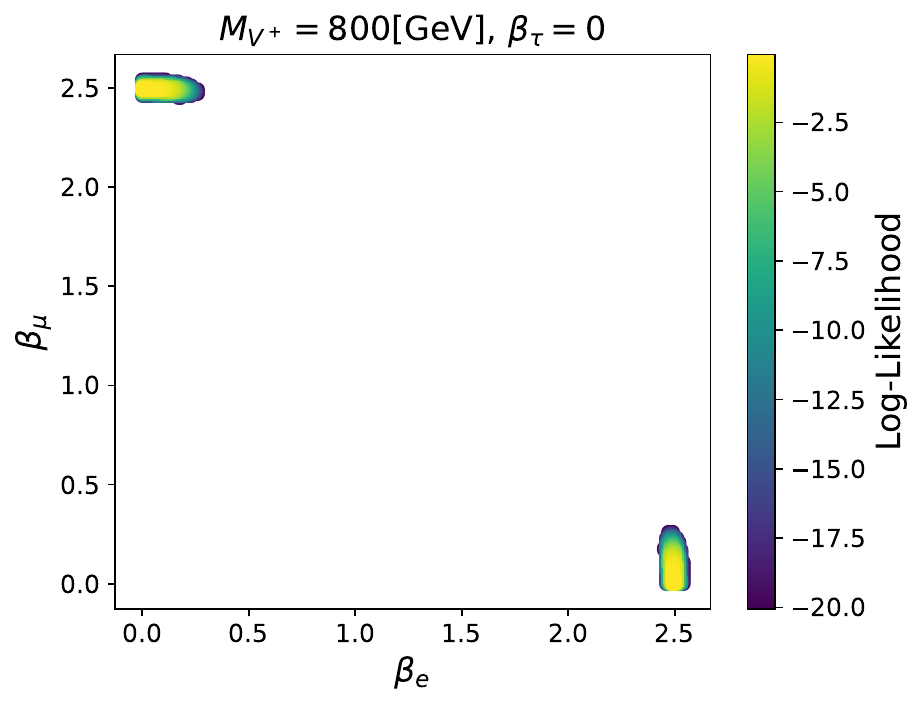}
            \caption{}
    \end{subfigure}
    \begin{subfigure}{0.495\textwidth}
            \includegraphics[width=\textwidth]{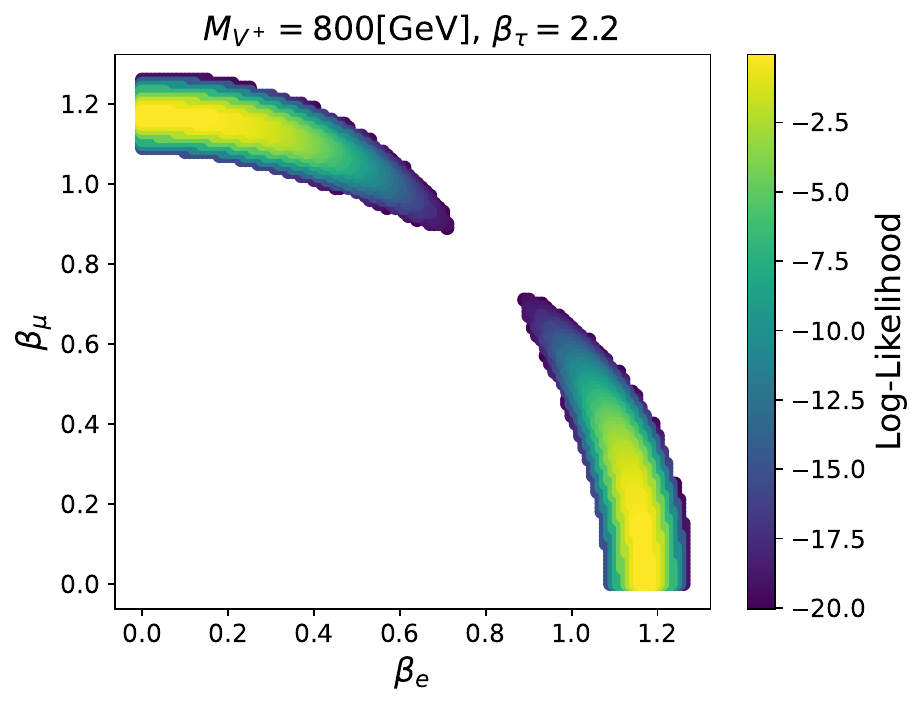}
            \caption{}
    \end{subfigure}
    
        \begin{subfigure}{0.495\textwidth}
            \includegraphics[width=\textwidth]{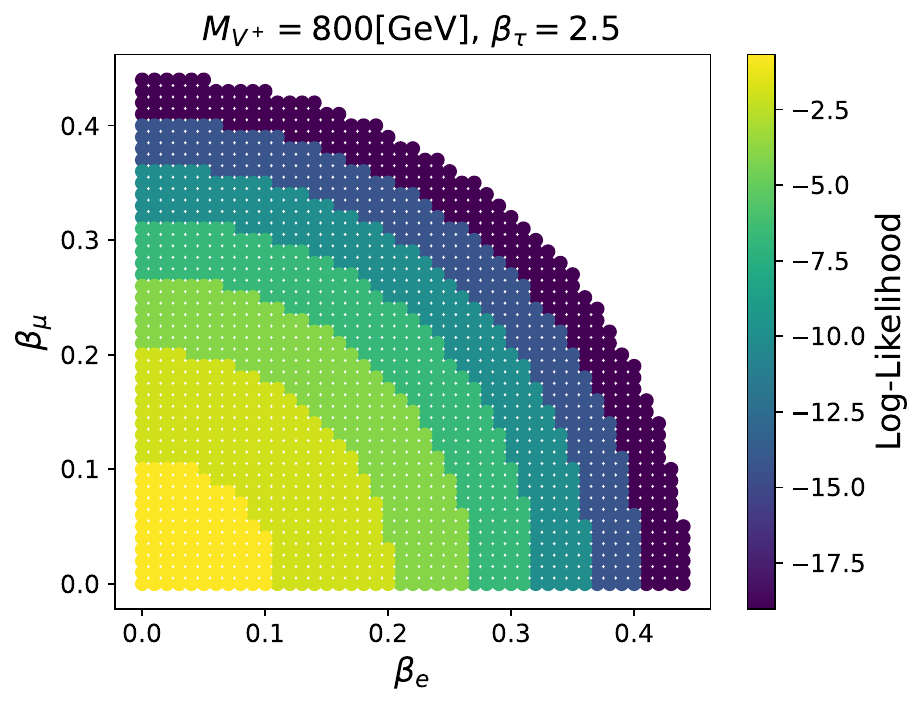}
            \caption{}
    \end{subfigure}
    \caption{Log-Likelihood profiles for some special cases, considering $M_N=50$[GeV]. The strong constraints arising from $\mu\to e\gamma$ impose that at least one coupling should smaller than the others.}
    \label{likelis}
\end{figure}

In order to find points in the parameter space that satisfy both DM and LFV constraint, we implemented a Log-Likelihood function:
\begin{equation}
    \ln \mathcal{L}=\ln \mathcal{L}^{DM}+\ln \mathcal{L}^{\mu\to e\gamma}+ \ln \mathcal{L}^{\tau\to \mu \gamma}+\ln \mathcal{L}^{\tau\to e \gamma},
\end{equation}
where all the likelihood functions are Gaussian. For the dark matter, we have centered the Gaussian on the Planck measurement for relic abundance, and we have set the standard deviation equal to the experimental uncertainty. For the LFV constraint, we have considered Gaussian likelihoods centered at 0 with a standard deviation equal to the upper bounds presented in Table \ref{lfv_pdg} (in concordance with the method presented in Ref. \cite{scotosinglet}). Some representative likelihood profiles are shown in Figure \ref{likelis}. The scenario with $\beta_e\sim \beta_\mu\sim \beta_\tau$ is practically excluded, however, the maximum likelihood is obtained when the dark fermion couples to one lepton family only. The points satisfying $\beta_i=0$ and $\beta_j<<\beta_k$ are very close to the maximum likelihood\footnote{The  Log-likelihood difference, $\Delta=\ln{\mathcal{L}}^{max}-\ln{\mathcal{L}}$ for these points is proportional to $10^{-13}$.} for any combination of $i,j,k$

\section{Indirect detection}\label{sec:indirect}
\begin{figure}[!h]
    \centering
    \begin{subfigure}{0.495\textwidth}
            \includegraphics[width=\textwidth]{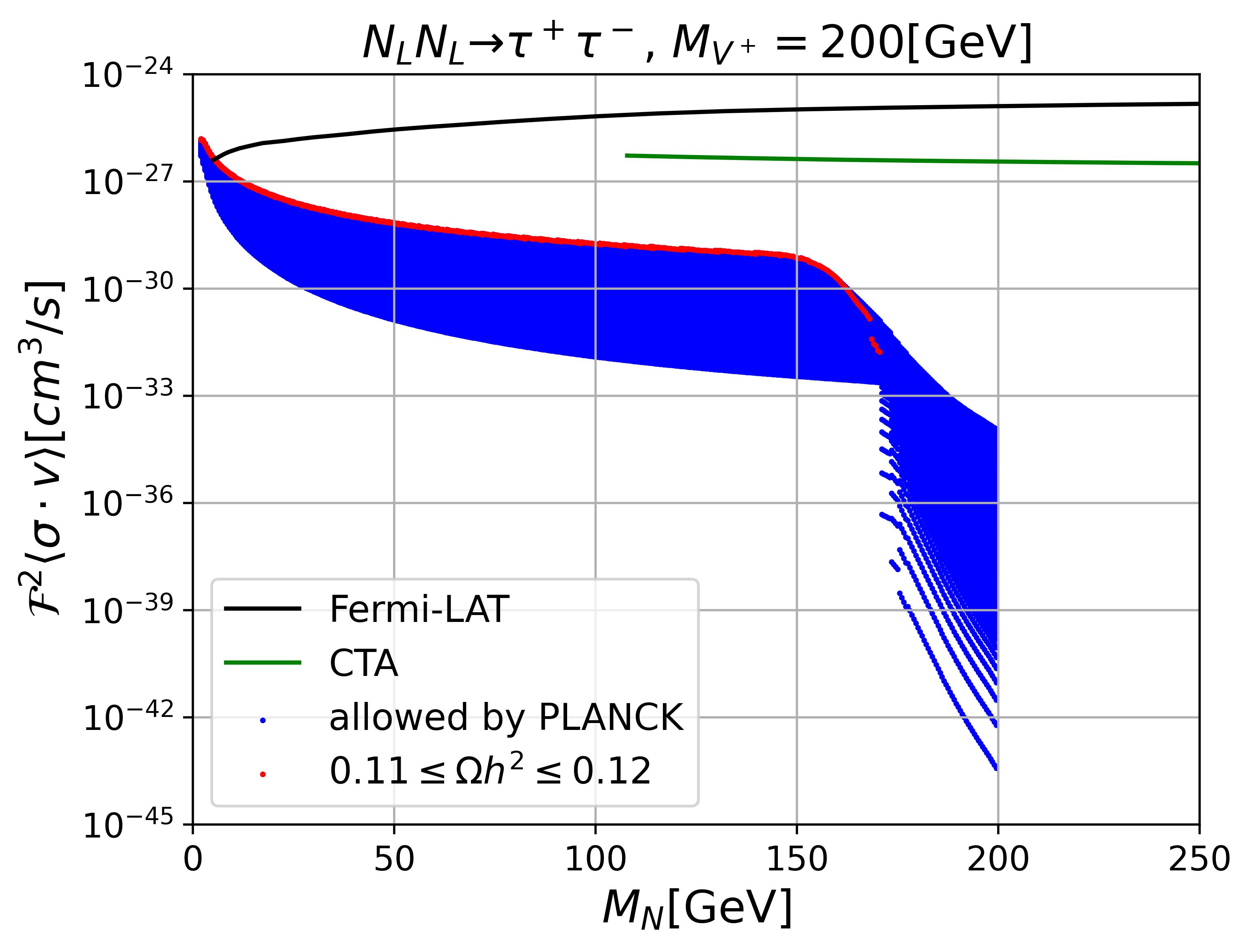}
            \caption{}
    \end{subfigure}
    \begin{subfigure}{0.495\textwidth}
            \includegraphics[width=\textwidth]{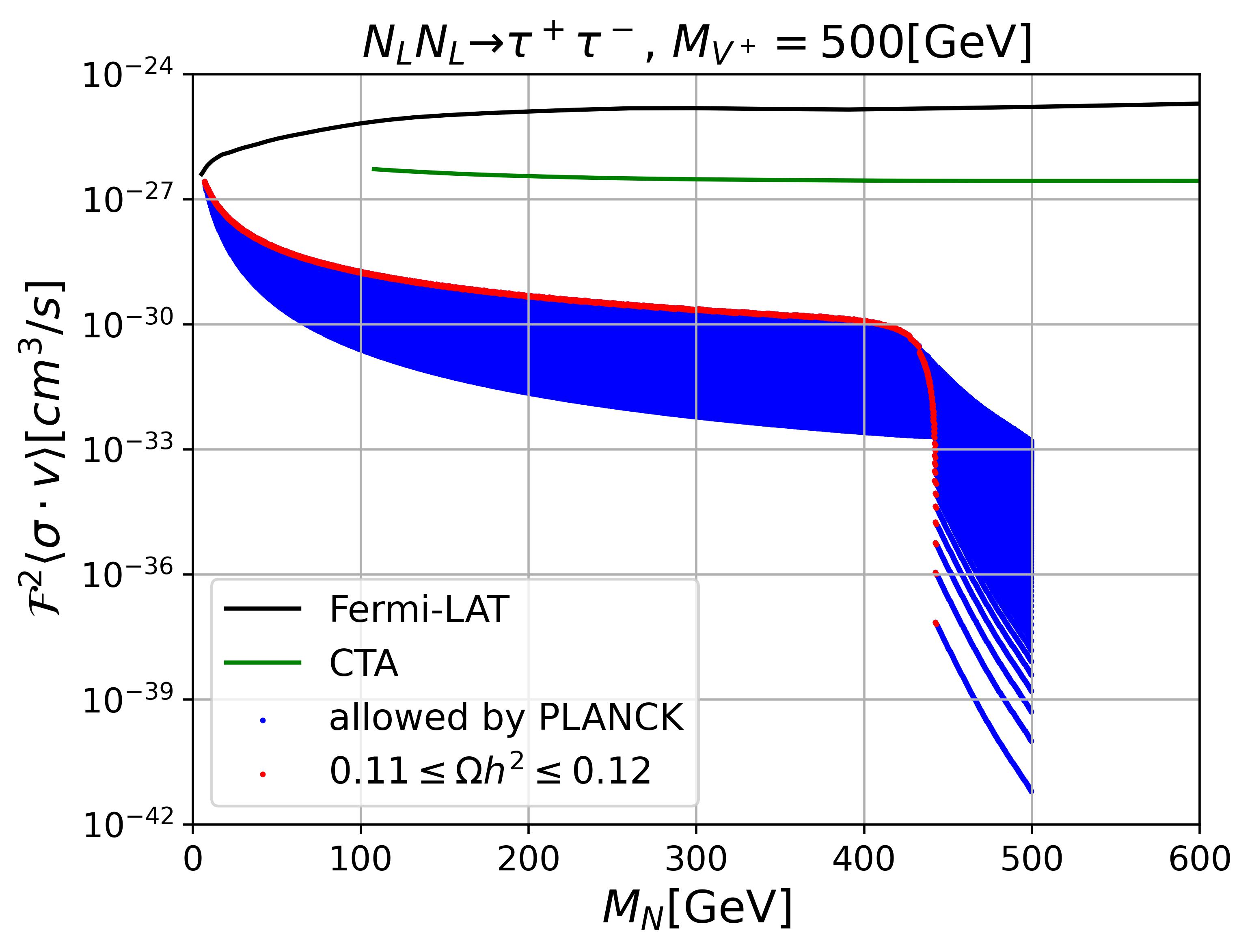}
            \caption{}
    \end{subfigure}
        \begin{subfigure}{0.495\textwidth}
            \includegraphics[width=\textwidth]{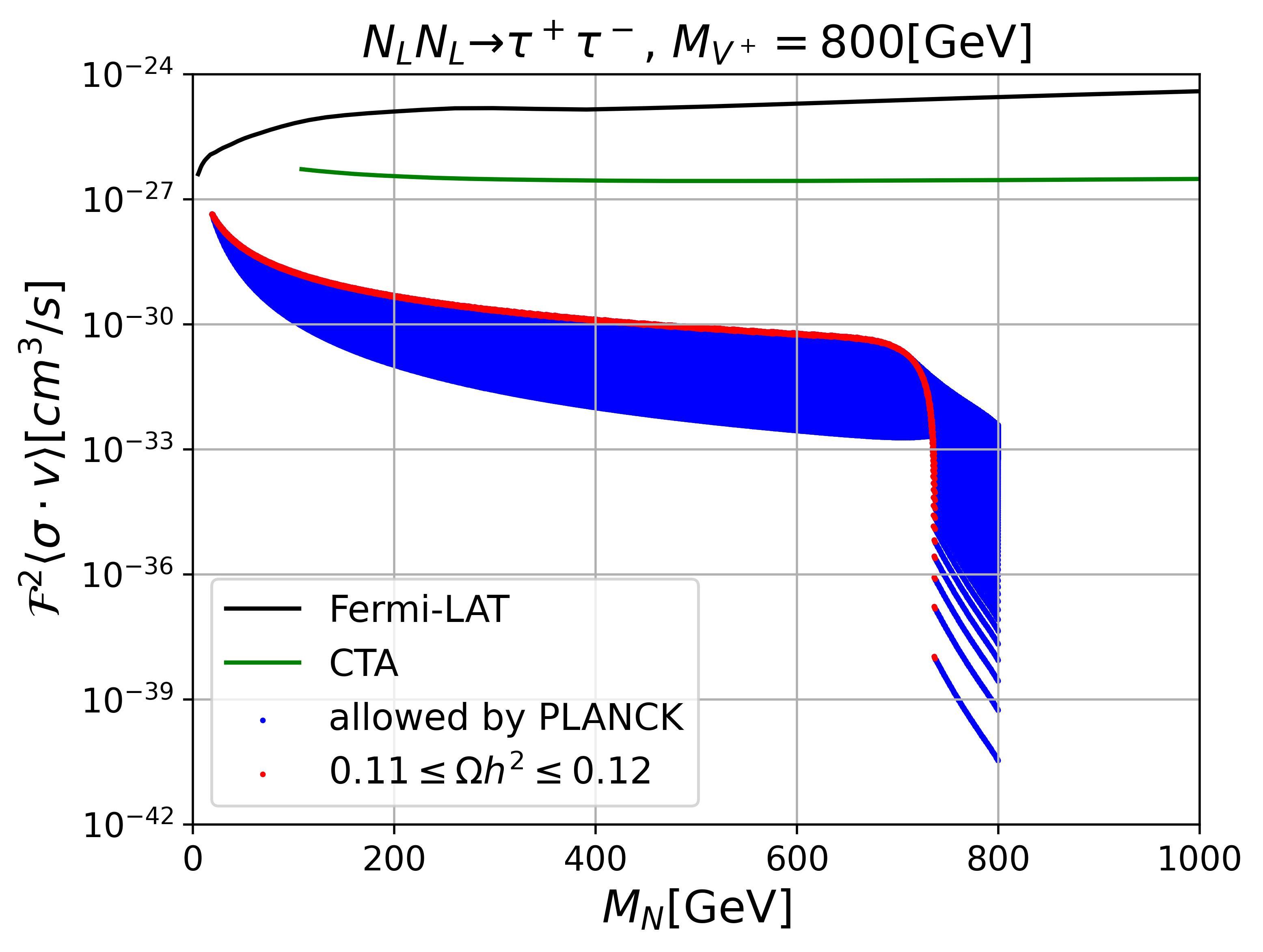}
            \caption{}
    \end{subfigure}
    \caption{Model prediction for indirect detection. It's worth recalling that these plots were obtained considering $\beta_e=\beta_\mu=0$. This choice gives the most optimistic prediction because higher values for these parameters would reduce the weighting factor.}
    \label{indirect}
\end{figure}

The annihilation cross section into charged leptons has the following form:
\begin{equation}
    \langle \sigma v\rangle(N_LN_L\to l^+l^-)= \sqrt{1-\frac{m_l^2}{M_N^2}}\frac{m_l^2|\beta_l|^4(2M_{V^+}^2+M_N^2-m_l^2)}{64\pi M_{V^+}^4(M_{V^+}^2+M_N^2-m_l^2)}.
    \label{ancs}
\end{equation}
From Eq.~\ref{ancs}, it can be seen that the most promising channel for indirect detection is $N_LN_L\to \tau^+ \tau^-$ since $\langle \sigma v\rangle$ as a manifestation of helicity suppression. Therefore, we have studied this signal considering the exclusion limits from the Fermi-LAT 
 telescope \cite{fermilims}. On the other hand, the expected sensitivity of CTA \cite{ctalims} could reach the threshold for observing this process. It's worth mentioning that these limits were obtained by assuming that DM relic density is saturated by a single type of particle, therefore we defined $\mathcal{F}=\frac{\Omega h^2}{0.12}$ as a weight for underabundant parameter space points. Since annihilation implies the interaction between two dark matter particles, the annihilation cross section must be weighted as $\mathcal{F}^2\langle \sigma v \rangle$. The result can is shown in Figure \ref{indirect}.
In general, the model prediction is much lower than the expected sensitivity of CTA, making the model hard to probe in the near future by means of CTA. However, there is a small region of the parameter space for low values of $M_N$ and $M_{V^+}$ that is excluded by Fermi-LAT,

\section{Direct detection}\label{sec:direct}
\begin{figure}[!h]
    \centering
    \begin{subfigure}{0.495\textwidth}
            \includegraphics[width=\textwidth]{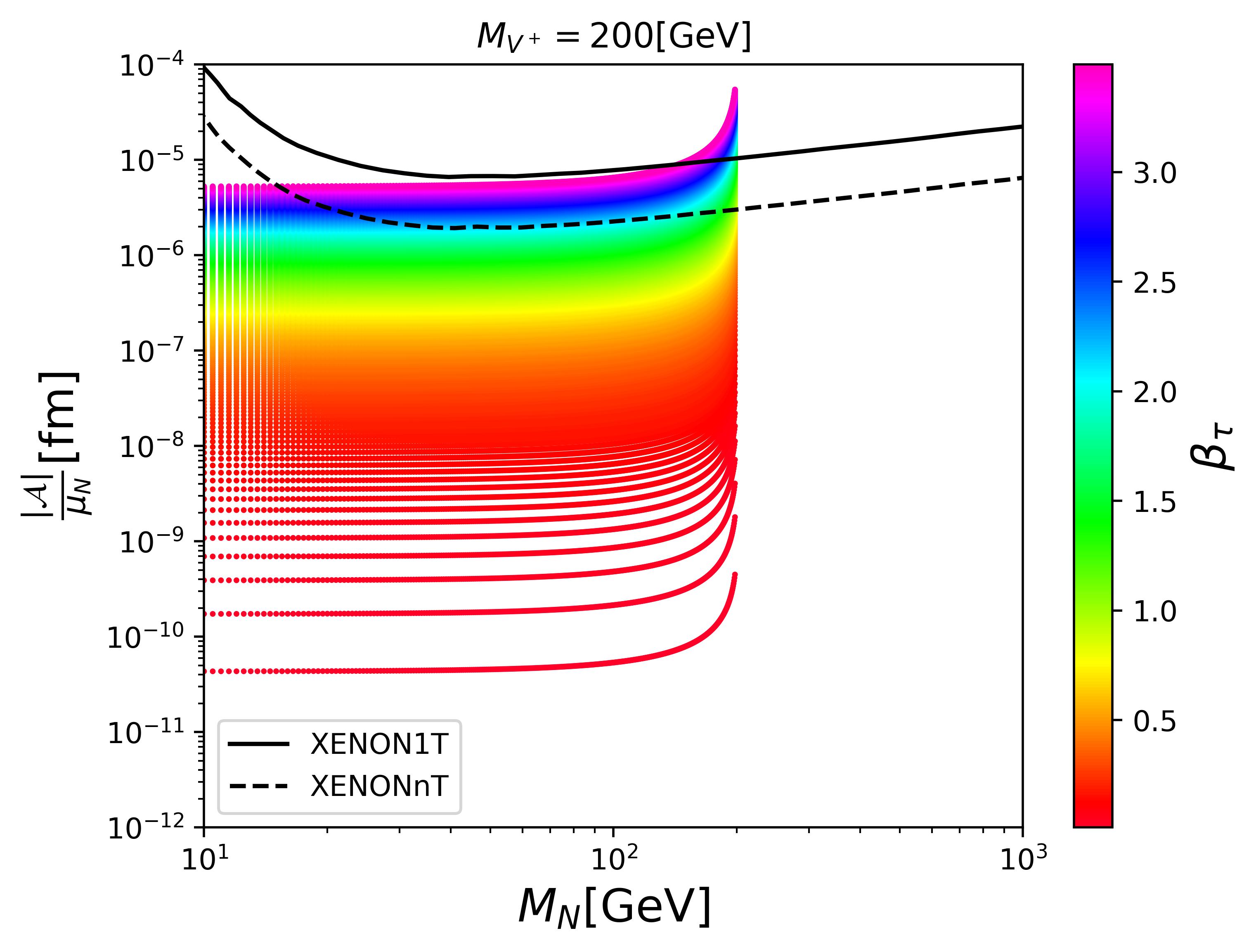}
            \caption{}
    \end{subfigure}
    \begin{subfigure}{0.495\textwidth}
            \includegraphics[width=\textwidth]{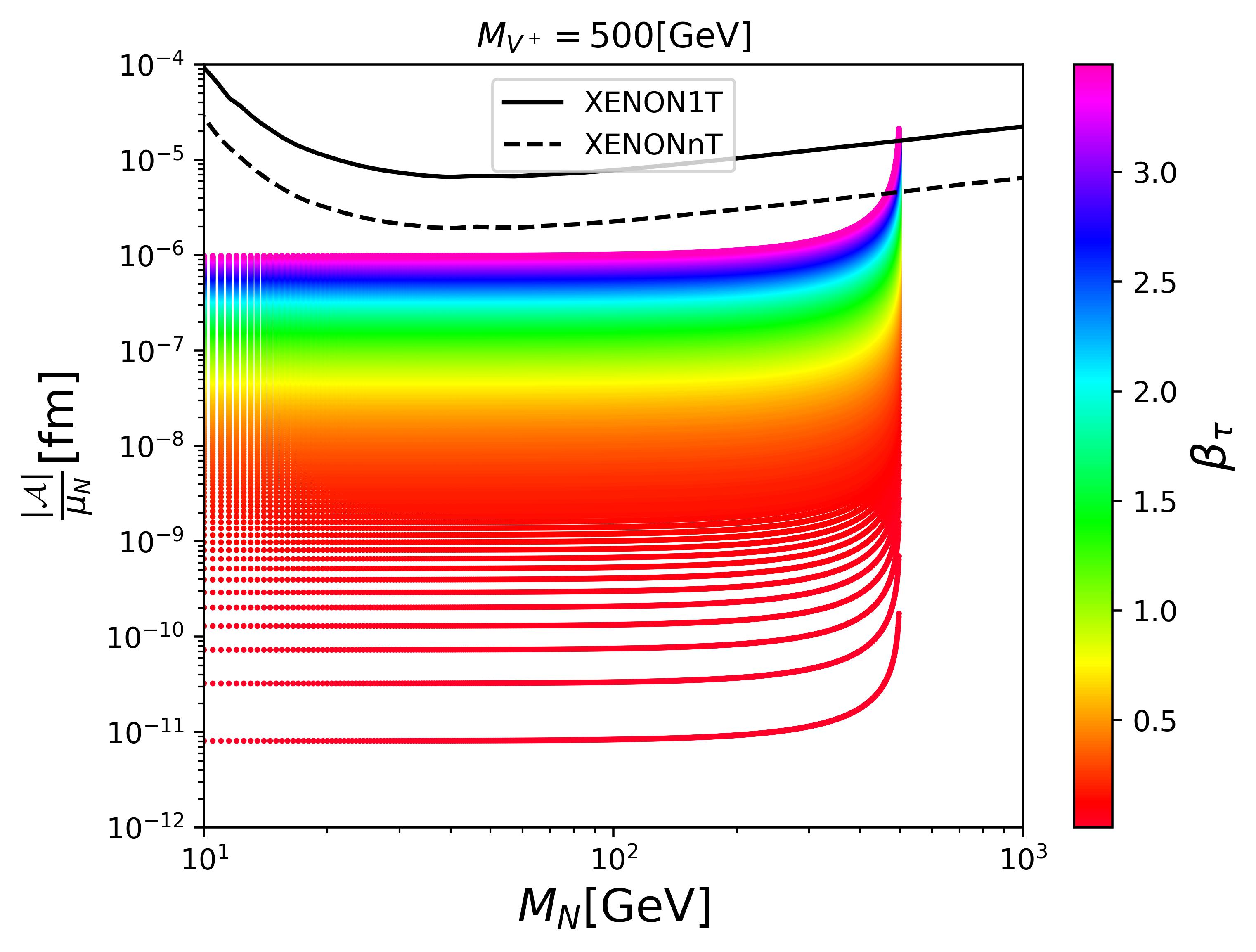}
            \caption{}
    \end{subfigure}
        \begin{subfigure}{0.495\textwidth}
            \includegraphics[width=\textwidth]{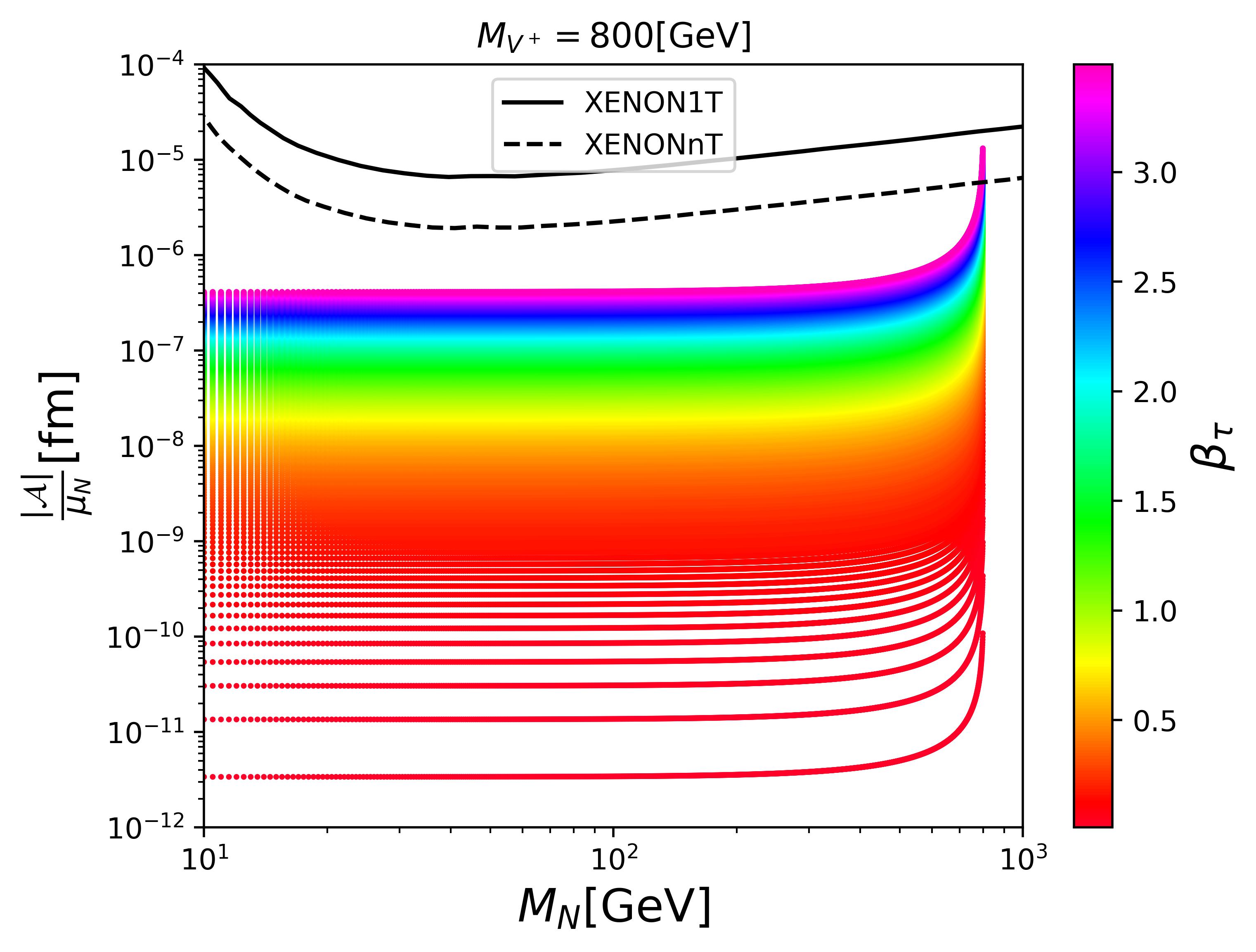}
            \caption{}
    \end{subfigure}
    \caption{Model prediction for direct detection. It's worth recalling that these plots were obtained considering $\beta_e=\beta_\mu=0$. The anapole moment is normalized by the nuclear magnetron $\mu_N$ defined as  $\mu_N=e/2m_p$}
    \label{direct}
\end{figure}

\begin{figure}[!h]
    \centering
    \includegraphics[width=0.3\textwidth]{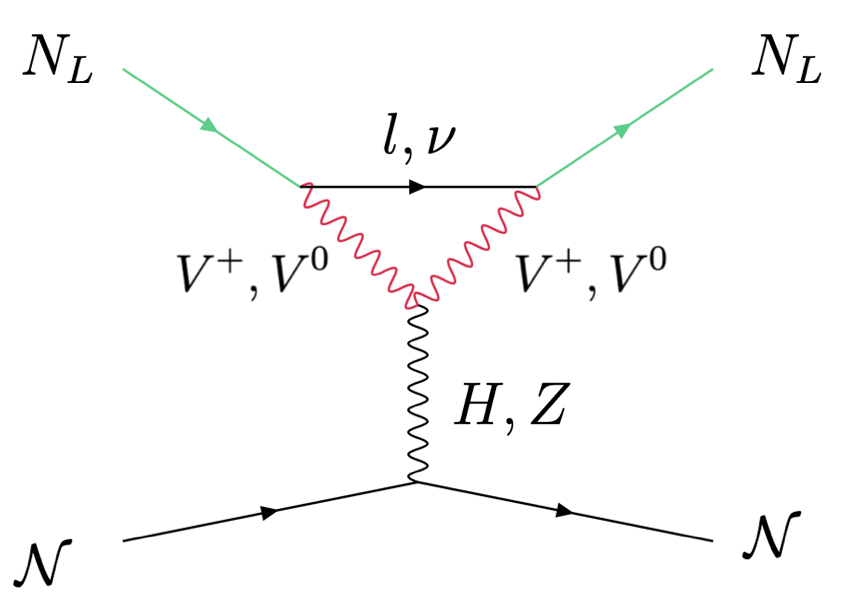}
    \caption{One loop contribution to DM-nucleon cross section considering fermion dark matter. It's worth mentioning that there is another contribution coming from the coupling of electroweak bosons to leptons, however the topology is quite similar.}
    \label{directdetection_diagram}
\end{figure}

Since we restricted ourselves to the physics case of fermion dark matter, DM interactions with nucleons arise only at 1 loop level (See Figure \ref{directdetection_diagram}), as expected for socotogenic-like models \cite{Schmidt:2012yg}. The main contribution comes from the anomalous coupling of majorana fermions to the photon, which is known as the anapole moment \cite{Ibarra:2022nzm}. This interaction is described by the effective interaction:
\begin{equation}
    \mathcal{L}_{eff}=\frac{\mathcal{A}}{2}\bar{N}_L \gamma_\mu \gamma_5 N_L\partial_\nu F^{\mu\nu},
\end{equation}
where $F^{\mu\nu}$ is the fiels strength of the photon. According to Ref. \cite{Ibarra:2022nzm}, the parameter $\mathcal{A}$ gas the following form:
\begin{equation}
    \mathcal{A}=\sum_{i=e,\mu,\tau} \frac{2e}{96\pi^2 M_N^2}|\beta_i|^2\mathcal{F}\left(\frac{m_i}{M_N},\frac{M_{V^+}}{M_N}\right),
\end{equation}
where
\begin{equation}
    \mathcal{F}(\mu,\eta)=\frac{3}{2}\log(\mu^2/\eta^2)+(3\eta^2-3\mu^2-7)f(\mu,\eta),
\end{equation}
and
\begin{equation}
    f(\mu,\eta)=\begin{cases}
     \frac{1}{2\sqrt{\Delta}}\log\left(\frac{\mu^2+\eta^2-1+\sqrt{\Delta}}{\mu^2+\eta^2-1-\sqrt{\Delta}}\right)   \quad &\text{if} \quad \Delta \neq 0 \\
     \frac{2}{(\mu^2-\eta^2)^2-1}\quad &\text{if} \quad \Delta=0
    \end{cases}
\end{equation}
with $\Delta \equiv (\mu^2+\eta^2-1)^2-4\mu^2\eta^2$. Note that this result is valid on the assumption of $\kappa=-1$, where the charged vector interaction with the photon has the same structure than the W. As is shown in Ref. \cite{Ibarra:2022nzm}, the anapole moment is constrained by direct detection experiments, in particular, we focused our attention on exclusion limits provided by XENON1T \cite{XENON:2018voc}, as long with the expected sensitivity expected at XENONnT \cite{XENON:2015gkh}. As can be seen in Figure \ref{direct}, most of parameter space points avoid these constraints, however the kinematical regime with small vector masses and large couplings could be probed in the near future.

 \section{Vector production at the LHC}\label{sec:vector_production}
\begin{figure}[!h]
    \centering
    \includegraphics[width=0.6\textwidth]{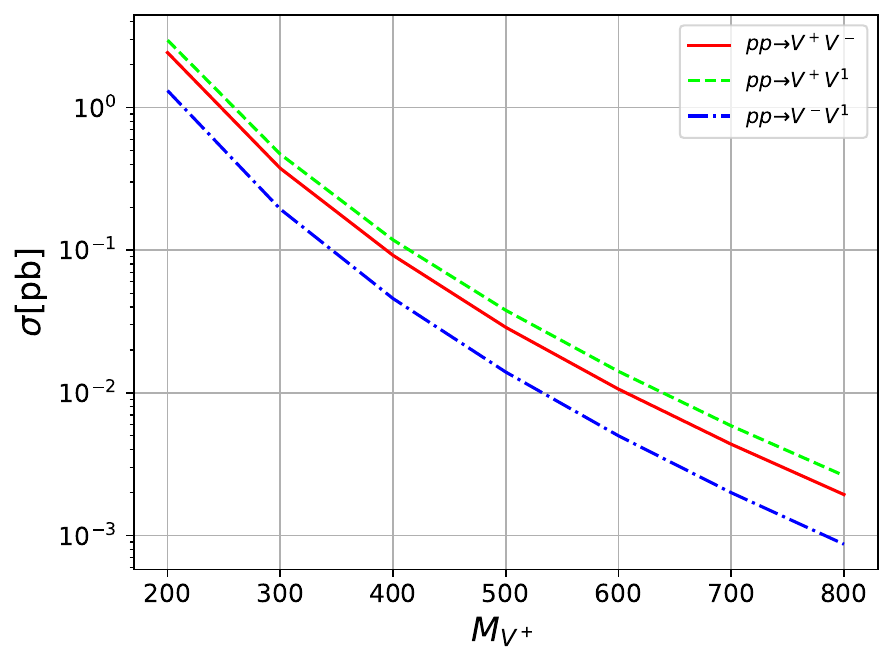}
    \caption{Parton level cross sections for the production of the vector states at the LHC, considering $\sqrt{s}=13$[TeV].}
    \label{fig:cross_vv}
\end{figure}

The structure of the model allows the electroweak pair production of the new vector states at the LHC. We simulated the parton level production of vector states using Madgraph5 aMC@NLO version 3.5.0 \cite{mg5} focusing on the scenario where at least one of the vectors is charged, this is because the neutral components decay only via $V^1\to N_L \nu$, which doesn't leave detector traces, therefore the production of neutral vectors will produce the same signal studied in Ref. \cite{vector_dm} regarding mono-X signals and shall not be considered in the present work. Results can be seen in Figure \ref{fig:cross_vv}.

\begin{figure}
    \centering
    \includegraphics[width=0.6\textwidth]{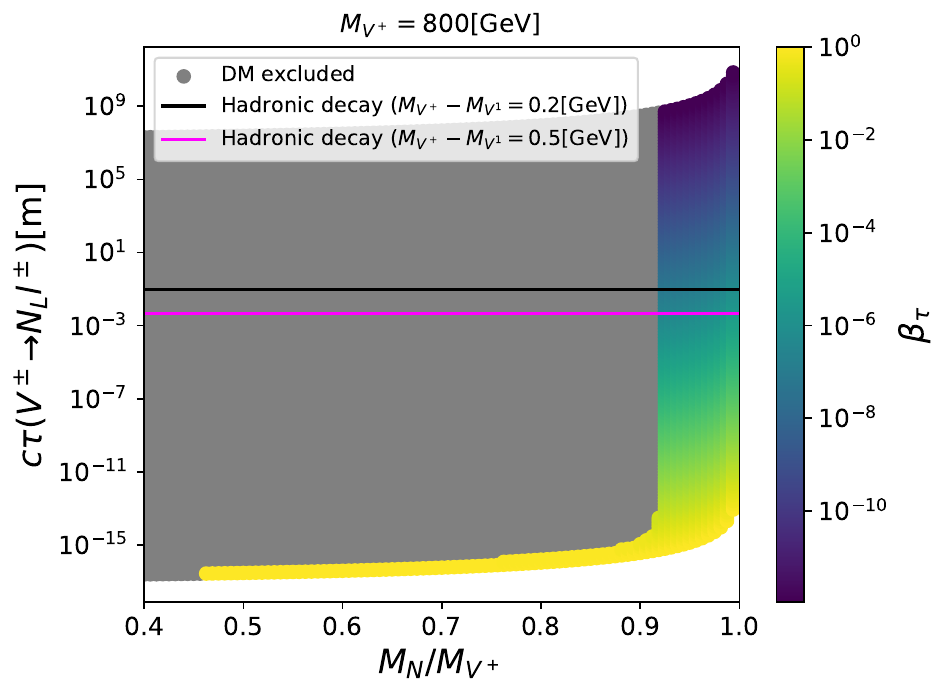}
    \caption{Decay length as a function of $M_N/M_{V^+}$. For this plot, we considered $\beta_e=\beta_\mu=0$. In this plot it's clear that, in the strong coannihilation regime, the $\beta$ plays a secondary role on achieving the correct relic density, allowing for lower values for this parameter.}
    \label{llp}
\end{figure}

\begin{figure}
    \centering
    \includegraphics[width=0.6\textwidth]{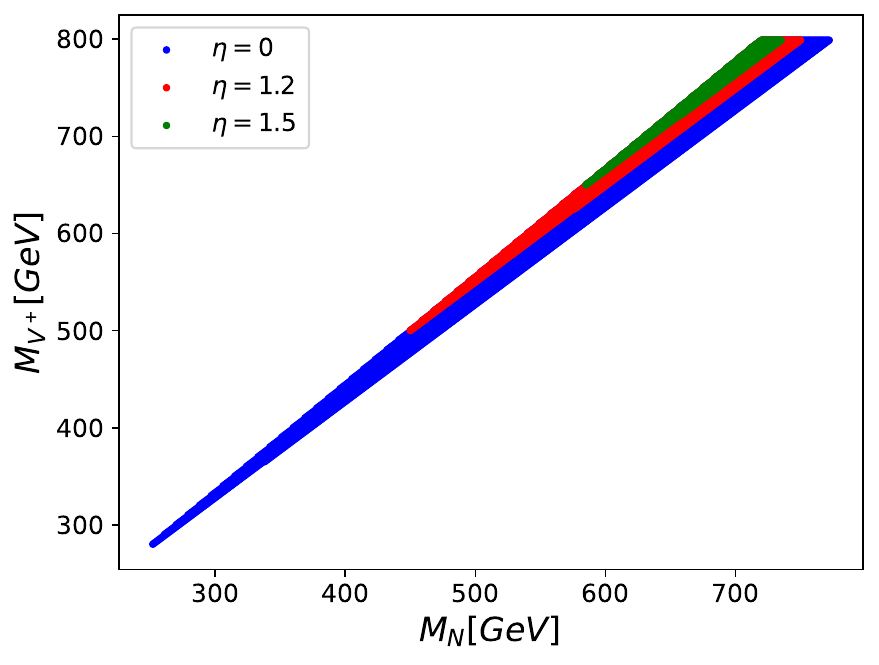}
    \caption{Region of the parameter space that can be probed  by means of displaced muons searches. Since these searches are sensitive to the muon $p_T$, the sensitive region is highly dependent on the kinematics of the event}. These limits are derived 
 from neglecting the muon mass.
    \label{llp_params}
\end{figure}

Of course, the vector states are unstable in the fermion dark matter scenario, therefore we need to perform a detailed study of the charged vector decay width. The main contribution comes from $V^\pm\to N_L l^\pm$:

\begin{equation}
        \Gamma(V^\pm\to N_L l^\pm)=(|\beta_e|^2 + |\beta_\mu|^2+|\beta_\tau|^2)\frac{ (M_{V^+}^2 - M_N^2)^2(2 M_{V^+}^2 +M_N^2) }{(48 \pi M_{V^+}^5)}.
\end{equation}
However, the hadronic decay $V^{\pm} \to V^{1,2} \pi^\pm$ may become relevant if we relax our assumption about the mass splitting of the vector states. Following the procedure of \cite{triplet_dm}, we introduce the W-pion mixing
\begin{equation}
    \mathcal{L}_{W\pi}=\frac{g f_\pi V_{ud}}{2\sqrt{2}}W_\mu^+\partial^\mu \pi^- +\text{h.c.},
\end{equation}
where $f_{\pi}=130$[MeV] is the pion decay constant and $V_{ud}$ is the element of the CKM matrix regarding the pion constituents.. Under this setup, the decay width has the following form:
\begin{equation}
\begin{split}
\Gamma(V^\pm\to V^1\pi^\pm)=&\frac{g_w^4f_\pi^2V_{ud}^2}{6144\pi M_W^4 M_{V^+}^5M_{V^1}^2}(M_{V^1}^4+(M_{V^+}^2-m_\pi^2)^2+2M_{V^+}^2(5M_{V^+}^2-m_\pi^2))\times\\
&(M_{V^+}^2-M_{V^1}^2)^2 
\sqrt{m_\pi^4-2m_\pi^2(M_{V^+}^2+M_{V^1}^2)+(M_{V^+}^2-M_{V^1}^2)^2}.
\end{split}
\end{equation}
Even if we would have considered a mass splitting between the vector fields, the leptonic decay is larger than the hadronic. As can be seen in Figure \ref{llp}, the leptonic decay is considerably larger, except on the region where $M_N \sim M_{V^+}$ and $\beta_i\leq 10^{-5}$. Besides, the pion decay only opens at $M_{V^+ -M_{V^1}}\geq 134$[MeV]. For lower values the vector mass splitting, the leptonic decay dominates. We focused our attention in the region where $M_N\sim M_{V^+}$, since the $\beta$ couplings can be small and avoid dark matter over abundance. This region could be probed at CMS \cite{CMS-PAS-EXO-14-012}. According to this reference, displaced muons   with $p_T >26$[GeV] can be reconstructed (there are additional cuts for the definition of the acceptance region, but for the sake of a general discussion we will consider only this cut). The energy of the displaced muon depends only on the kinematics and therefore on the mass difference between the vector and the dark fermion. Taking this considerations, this region of the parameter space can be probed under this setup, as can be seen in Figure \ref{llp_params}.

\section{ PROMPT SEARCHES AT THE LHC}\label{sec:collider}

In a previous work, we demonstrated that the production of 2 dark fermions can produce a distinctive signal at the LHC \cite{colliderLHNL}, which is composed by a same flavor opposite sign lepton pair and missing energy. In that work, we fixed $\kappa=1$, therefore we used Madgraph5 aMC@NLO version 3.5.0 \cite{mg5}  to compute the cross section of the process $pp\to N_L N_L \mu^+ \mu^-$ for some benchmark points and $\kappa=1,-1$, as can be seen in Table \ref{likelimax}. Firstly, we can note that the cross sections are slightly higher for $\kappa=-1$ for all benchmark points. On the other hand,
we notice that the scenario where $\beta_\mu>>\beta_e,\beta_\tau $ predicts cross sections that are inconsistent with the ATLAS upper limits \cite{susylims_dy}, therefore this scenario is disregarded. Finally, the scenario where $\beta_\tau>>\beta_e,\beta_\mu $ predicts smaller cross sections due to the decay width suppression, making this scenario more promising, predicting small cross sections at colliders and the highest possible indirect detection rate. 
\begin{table}[!h]
    \centering
    \begin{tabular}{|c|c|c|c|c|c|}
    \hline
       $M_{V^+}$[GeV] & $\beta_e$ & $\beta_\mu$ &$\beta_\tau$ &$\sigma_{\kappa=1}$[fb]&$\sigma_{\kappa=-1}$[fb]  \\
       \hline
                $200$     & $0.00$ & $0.06$ &$0.63$&$6.98\times 10^{-5}$&$1.29\times 10^{-4}$\\   
                $200$     & $0.00$ & $0.64$ &$0.00$&$1.1\times 10^{3}$&$2\times 10^{3}$\\   
                $500$     & $0.00$ &  $0.01$ &$1.56$&$1.47\times 10^{-7}$&$4.21\times 10^{-8}$\\
                $500$     & $0.00$ &  $1.56$ &$0.00$&$8.7$&$24.9$\\        
                $800$     & $0.00$ &  $0.06$ &$2.49$&$1.56\times 10^{-7}$&$5.71\times 10^{-7}$\\
                $800$     & $0.00$ &  $2.5$ &$0.00$&$0.47$&$1.7$\\
                \hline
    \end{tabular}
    \caption{Benchmark points that saturate relic density and the parton level production cross section for different values of $\kappa$. All the cross sections were obtained for a fixed value of $M_N=50$[GeV].}
    \label{likelimax}
\end{table}
For the sake of completeness, we computed the production cross section for different parameter space points, relating this value with the predicted relic density. 
following the methodology presented in our previous work \cite{colliderLHNL}, we corrected the parton level cross section by means of considering detector efficiency effects and relevant selection criteria, obtaining:
\begin{equation}
    \sigma_{eff}=\epsilon \mathcal{A} \sigma_{parton},
\end{equation}
where $\epsilon$ parametrizes the detector efficiency and $\mathcal{A}$ is the acceptance obtained from applying the cutflow (more details in \cite{colliderLHNL}). With this effective cross section, we are able to recast ATLAS results \cite{susylims_dy} for searches in the dilepton + $\slashed{E}_T$ channel. It's important to note that, ince we are interested in HL-LHC projections, we are keeping the most optimistic value of $\epsilon=0.55 $ for the detector efficiency.
 
\begin{figure}[!h]
    \begin{subfigure}{0.6\textwidth}
            \includegraphics[width=\textwidth]{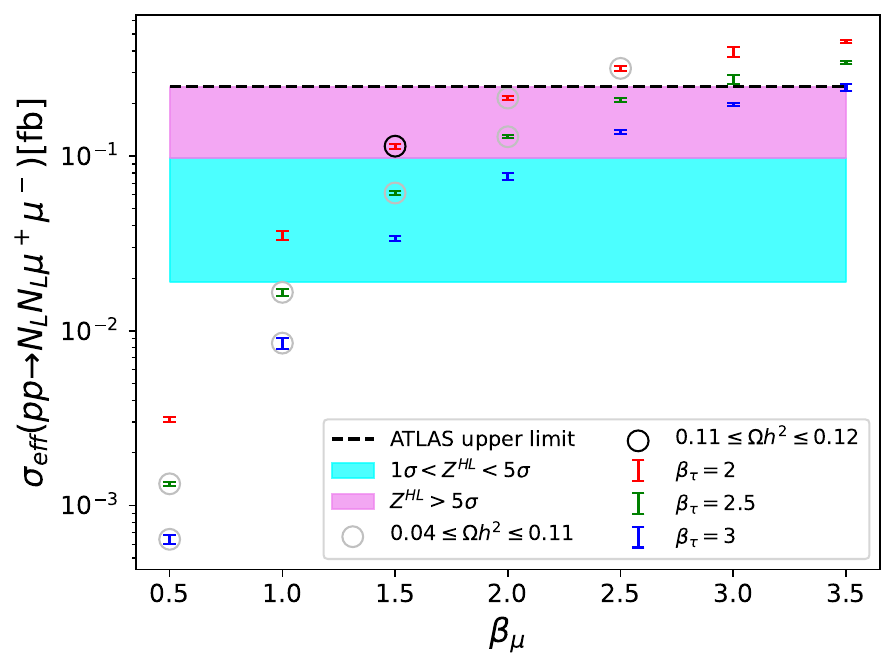}
            \caption{}
    \end{subfigure}

    \begin{subfigure}{0.6\textwidth}
            \includegraphics[width=\textwidth]{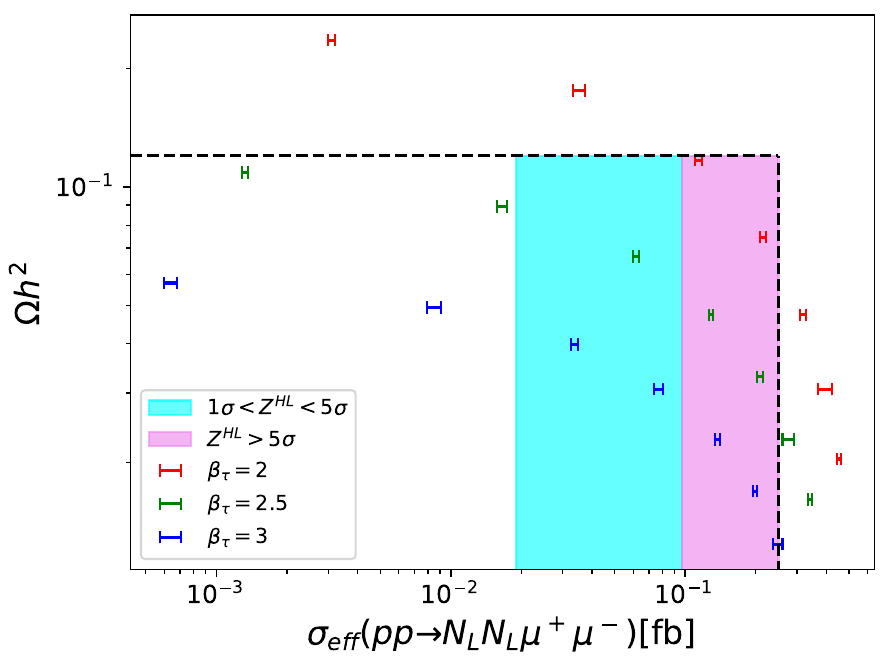}
            \caption{}
    \end{subfigure}
    \caption{Correlations between the production cross section and relic density, for $\beta_e=0,\kappa=-1,M_{V^+}=800\text{[GeV]}, M_N=50\text{[GeV]}$ and different values for $\beta_\mu$. The dashed lines represent the exclusion limits in both observables.}
    \label{cross_vs_relic}
\end{figure}

As can be seen in Figure \ref{cross_vs_relic}, there is an inverse relation between these two variables, and the $\beta_\tau$ parameter plays a key role in this correlation. Additionally, we included the expected sensitivities $Z^{HL}$ for the HL-LHC at $3000[\text{fb}^{-1}]$, showing that the model can account for a significant fraction of the dark matter relic abundance and be probed at the HL-LHC.

\section{Conclusions}\label{sec:conclusions}

In this work, we studied the dark matter phenomenology of the Vector Scotogenic Model, composed by a massive vector doublet under $SU(2)_L$ and a left-handed  Majorana fermion. We focused on the scenario where the fermion is the dark matter candidate. We separated the parameter space into two regions, depending on the mass splitting between the singlet fermion and the new vector states. When the fermion mass is not comparable to the vector mass, the early universe dynamics is dominated by fermion annihilation, and it is possible to define lower bounds on the couplings between the new fields and the SM leptons. When the mass split is small enough, vectors have a significant contribution to the annihilation cross section, contributing to the relic density even when the above mentioned couplings are small. On the other hand, both regimes can be probed at colliders, the first one with prompt searches and the second one under the paradigm of long living particles. Besides that, the annihilation cross section is too small to be probed in the context of indirect detection, however, the anomalous coupling of the Majorana fermion to photons via the anapole moment allows us to make optimistic predictions for direct detection facilities.

\section*{Aknowledgements}
This work was funded by ANID - Millennium Program - ICN2019\_044. Also, we would like to thank to the DGIIP-UTFSM for funding during the development of this work. AZ was partially supported
by Proyecto ANID PIA/APOYO AFB220004 (Chile) and Fondecyt 1230110. Also, JZ was partially supported by Fondecyt 1240216. Finally, PAC is grateful to Toshihiko Ota for insightful discussions related with particle cosmology in the early universe.

\section{Appendix A: Comparison of the complete and approximated calculation of relic density}\label{sec:ap_2}
\begin{figure}[!h]
    \centering
    \includegraphics[width=0.6\textwidth]{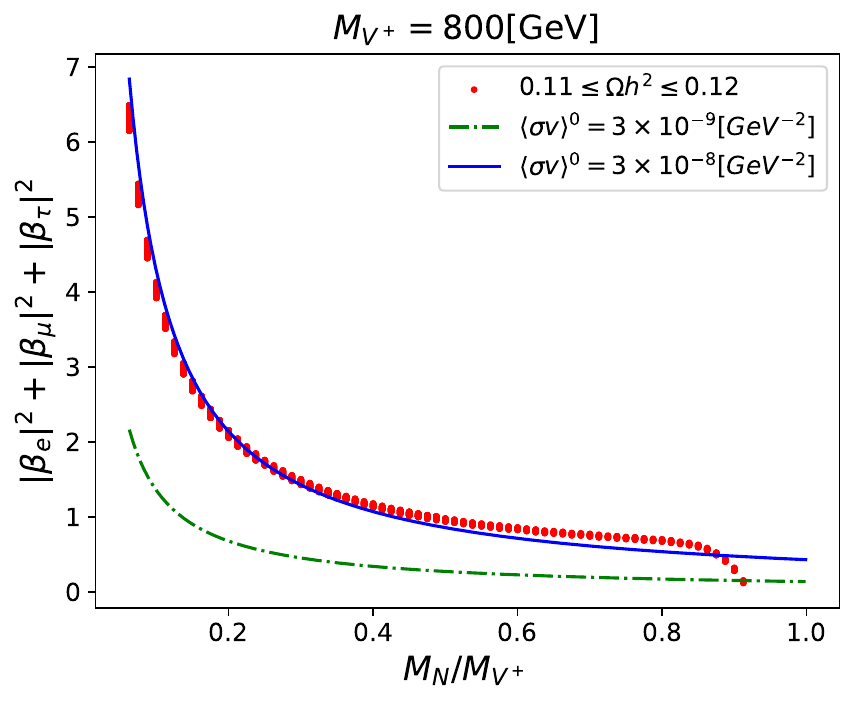}
    \caption{Comparison between the rough approximation considered in our previous work and the relic density computation with micrOMEGAs.}
    \label{comparison}
\end{figure}

In our previous work, we used the results from Ref. \cite{dong2021} to set lower limits on the parameter space, using the following expression:
    \begin{equation}\label{eq:viet}
    \langle \sigma v\rangle^0=\sum_{k,k'=\{e,\mu,\tau\}}|\beta_k^*\beta_{k'}|^2\frac{M_N^2}{8\pi}\left(1+\frac{8T_f}{M_N}\right)\left(\frac{1}{M_{V^+}^4}+\frac{4}{(M_{V^1}^2+M_{V^2}^2)^2}\right).
\end{equation}
 This quantity allows to estimate dark matter relic density, demanding that $\langle \sigma v\rangle^0 \geq 3\times 10^{-9}[GeV^{-2}]$. This constraint is a rough estimation, and therefore we decided to compare our present results with this expression. As can be seen in Figure \ref{comparison}, our current results present a stronger constraint on the parameter space. However, taking a more stringent limit of $\langle \sigma v\rangle^0 \geq 3\times 10^{-8}[GeV^{-2}]$ gives closer results to the micrOMEGAs calculation. On the other hand, Eq. \eqref{eq:viet} doesn't account for co-annihilations, therefore it fits well the region where $M_N<<M_{V^+}$, but it doesn't describe well the nearly degenerate regime.

\bibliographystyle{utphys}
\bibliography{References}

\providecommand{\href}[2]{#2}\begingroup\raggedright\begin{thebibliography}{10}

\bibitem{PLANCK}
{\bfseries Planck} Collaboration, N.~Aghanim {\em et~al.}, ``{Planck 2018
  results. VI. Cosmological parameters},''
  \href{http://dx.doi.org/10.1051/0004-6361/201833910}{{\em Astron. Astrophys.}
  {\bfseries 641} (2020) A6}, \href{http://arxiv.org/abs/1807.06209}{{\ttfamily
  arXiv:1807.06209 [astro-ph.CO]}}. [Erratum: Astron.Astrophys. 652, C4
  (2021)].

\bibitem{Ma:2006km}
E.~Ma, ``{Verifiable radiative seesaw mechanism of neutrino mass and dark
  matter},'' \href{http://dx.doi.org/10.1103/PhysRevD.73.077301}{{\em Phys.
  Rev. D} {\bfseries 73} (2006) 077301},
  \href{http://arxiv.org/abs/hep-ph/0601225}{{\ttfamily arXiv:hep-ph/0601225}}.

\bibitem{masses_and_mixings}
A.~E. C\'arcamo~Hern\'andez, J.~Vignatti, and A.~Zerwekh, ``{Generating lepton
  masses and mixings with a heavy vector doublet},''
  \href{http://dx.doi.org/10.1088/1361-6471/ab4499}{{\em J. Phys. G} {\bfseries
  46} no.~11, (2019) 115007}, \href{http://arxiv.org/abs/1807.05321}{{\ttfamily
  arXiv:1807.05321 [hep-ph]}}.

\bibitem{dong2021}
P.~Van~Dong, D.~Van~Loi, L.~D. Thien, and P.~N. Thu, ``{Novel imprint of a
  vector doublet},'' \href{http://dx.doi.org/10.1103/PhysRevD.104.035001}{{\em
  Phys. Rev. D} {\bfseries 104} no.~3, (2021) 035001},
  \href{http://arxiv.org/abs/2104.12160}{{\ttfamily arXiv:2104.12160
  [hep-ph]}}.

\bibitem{colliderLHNL}
P.~Areyuna~C., J.~Zamora-Saa, and A.~R. Zerwekh, ``{Probing left-handed heavy
  neutral leptons in the Vector Scotogenic Model},''
  \href{http://dx.doi.org/10.1007/JHEP02(2024)153}{{\em JHEP} {\bfseries 02}
  (2024) 153}, \href{http://arxiv.org/abs/2211.09753}{{\ttfamily
  arXiv:2211.09753 [hep-ph]}}.

\bibitem{Tulin:2017ara}
S.~Tulin and H.-B. Yu, ``{Dark Matter Self-interactions and Small Scale
  Structure},'' \href{http://dx.doi.org/10.1016/j.physrep.2017.11.004}{{\em
  Phys. Rept.} {\bfseries 730} (2018) 1--57},
  \href{http://arxiv.org/abs/1705.02358}{{\ttfamily arXiv:1705.02358
  [hep-ph]}}.

\bibitem{vector_dm}
B.~D. S\'aez, F.~Rojas-Abatte, and A.~R. Zerwekh, ``{Dark Matter from a Vector
  Field in the Fundamental Representation of $SU(2)_L$},''
  \href{http://dx.doi.org/10.1103/PhysRevD.99.075026}{{\em Phys. Rev. D}
  {\bfseries 99} no.~7, (2019) 075026},
  \href{http://arxiv.org/abs/1810.06375}{{\ttfamily arXiv:1810.06375
  [hep-ph]}}.

\bibitem{micromegas1}
G.~Belanger, F.~Boudjema, A.~Pukhov, and A.~Semenov, ``{micrOMEGAs$\_$3: A
  program for calculating dark matter observables},''
  \href{http://dx.doi.org/10.1016/j.cpc.2013.10.016}{{\em Comput. Phys.
  Commun.} {\bfseries 185} (2014) 960--985},
  \href{http://arxiv.org/abs/1305.0237}{{\ttfamily arXiv:1305.0237 [hep-ph]}}.

\bibitem{micromegas3}
G.~Belanger, F.~Boudjema, A.~Pukhov, and A.~Semenov, ``{MicrOMEGAs 2.0: A
  Program to calculate the relic density of dark matter in a generic model},''
  \href{http://dx.doi.org/10.1016/j.cpc.2006.11.008}{{\em Comput. Phys.
  Commun.} {\bfseries 176} (2007) 367--382},
  \href{http://arxiv.org/abs/hep-ph/0607059}{{\ttfamily arXiv:hep-ph/0607059}}.

\bibitem{scalar_scoto}
C.~Hagedorn, J.~Herrero-Garc\'\i{}a, E.~Molinaro, and M.~A. Schmidt,
  ``{Phenomenology of the Generalised Scotogenic Model with Fermionic Dark
  Matter},'' \href{http://dx.doi.org/10.1007/JHEP11(2018)103}{{\em JHEP}
  {\bfseries 11} (2018) 103}, \href{http://arxiv.org/abs/1804.04117}{{\ttfamily
  arXiv:1804.04117 [hep-ph]}}.

\bibitem{Baumholzer:2019twf}
S.~Baumholzer, V.~Brdar, P.~Schwaller, and A.~Segner, ``{Shining Light on the
  Scotogenic Model: Interplay of Colliders and Cosmology},''
  \href{http://dx.doi.org/10.1007/JHEP09(2020)136}{{\em JHEP} {\bfseries 09}
  (2020) 136}, \href{http://arxiv.org/abs/1912.08215}{{\ttfamily
  arXiv:1912.08215 [hep-ph]}}.

\bibitem{freezeinfreezeout}
Y.~Du, F.~Huang, H.-L. Li, Y.-Z. Li, and J.-H. Yu, ``{Revisiting dark matter
  freeze-in and freeze-out through phase-space distribution},''
  \href{http://dx.doi.org/10.1088/1475-7516/2022/04/012}{{\em JCAP} {\bfseries
  04} no.~04, (2022) 012}, \href{http://arxiv.org/abs/2111.01267}{{\ttfamily
  arXiv:2111.01267 [hep-ph]}}.

\bibitem{ilakovac}
A.~Ilakovac and A.~Pilaftsis, ``{Flavor violating charged lepton decays in
  seesaw-type models},''
  \href{http://dx.doi.org/10.1016/0550-3213(94)00567-X}{{\em Nucl. Phys. B}
  {\bfseries 437} (1995) 491},
  \href{http://arxiv.org/abs/hep-ph/9403398}{{\ttfamily arXiv:hep-ph/9403398}}.

\bibitem{pdg2022}
{\bfseries Particle Data Group} Collaboration, R.~L. Workman {\em et~al.},
  ``{Review of Particle Physics},''
  \href{http://dx.doi.org/10.1093/ptep/ptac097}{{\em PTEP} {\bfseries 2022}
  (2022) 083C01}.

\bibitem{scotosinglet}
A.~Beniwal, J.~Herrero-Garc\'\i{}a, N.~Leerdam, M.~White, and A.~G. Williams,
  ``{The ScotoSinglet Model: a scalar singlet extension of the Scotogenic
  Model},'' \href{http://dx.doi.org/10.1007/JHEP06(2021)136}{{\em JHEP}
  {\bfseries 21} (2020) 136}, \href{http://arxiv.org/abs/2010.05937}{{\ttfamily
  arXiv:2010.05937 [hep-ph]}}.

\bibitem{fermilims}
M.~Di~Mauro, M.~Stref, and F.~Calore, ``{Investigating the effect of Milky~Way
  dwarf spheroidal galaxies extension on dark matter searches with Fermi-LAT
  data},'' \href{http://dx.doi.org/10.1103/PhysRevD.106.123032}{{\em Phys. Rev.
  D} {\bfseries 106} no.~12, (2022) 123032},
  \href{http://arxiv.org/abs/2212.06850}{{\ttfamily arXiv:2212.06850
  [astro-ph.HE]}}.

\bibitem{ctalims}
{\bfseries CTA} Collaboration, A.~Acharyya {\em et~al.}, ``{Sensitivity of the
  Cherenkov Telescope Array to a dark matter signal from the Galactic
  centre},'' \href{http://dx.doi.org/10.1088/1475-7516/2021/01/057}{{\em JCAP}
  {\bfseries 01} (2021) 057}, \href{http://arxiv.org/abs/2007.16129}{{\ttfamily
  arXiv:2007.16129 [astro-ph.HE]}}.

\bibitem{Schmidt:2012yg}
D.~Schmidt, T.~Schwetz, and T.~Toma, ``{Direct Detection of Leptophilic Dark
  Matter in a Model with Radiative Neutrino Masses},''
  \href{http://dx.doi.org/10.1103/PhysRevD.85.073009}{{\em Phys. Rev. D}
  {\bfseries 85} (2012) 073009},
  \href{http://arxiv.org/abs/1201.0906}{{\ttfamily arXiv:1201.0906 [hep-ph]}}.

\bibitem{Ibarra:2022nzm}
A.~Ibarra, M.~Reichard, and R.~Nagai, ``{Anapole moment of Majorana fermions
  and implications for direct detection of neutralino dark matter},''
  \href{http://dx.doi.org/10.1007/JHEP01(2023)086}{{\em JHEP} {\bfseries 01}
  (2023) 086}, \href{http://arxiv.org/abs/2207.01014}{{\ttfamily
  arXiv:2207.01014 [hep-ph]}}.

\bibitem{XENON:2018voc}
{\bfseries XENON} Collaboration, E.~Aprile {\em et~al.}, ``{Dark Matter Search
  Results from a One Ton-Year Exposure of XENON1T},''
  \href{http://dx.doi.org/10.1103/PhysRevLett.121.111302}{{\em Phys. Rev.
  Lett.} {\bfseries 121} no.~11, (2018) 111302},
  \href{http://arxiv.org/abs/1805.12562}{{\ttfamily arXiv:1805.12562
  [astro-ph.CO]}}.

\bibitem{XENON:2015gkh}
{\bfseries XENON} Collaboration, E.~Aprile {\em et~al.}, ``{Physics reach of
  the XENON1T dark matter experiment},''
  \href{http://dx.doi.org/10.1088/1475-7516/2016/04/027}{{\em JCAP} {\bfseries
  04} (2016) 027}, \href{http://arxiv.org/abs/1512.07501}{{\ttfamily
  arXiv:1512.07501 [physics.ins-det]}}.

\bibitem{mg5}
J.~Alwall, R.~Frederix, S.~Frixione, V.~Hirschi, F.~Maltoni, O.~Mattelaer,
  H.-S. Shao, T.~Stelzer, P.~Torrielli, and M.~Zaro, ``The automated
  computation of tree-level and next-to-leading order differential cross
  sections, and their matching to parton shower simulations,''
  \href{http://dx.doi.org/10.1007/jhep07(2014)079}{{\em Journal of High Energy
  Physics} {\bfseries 2014} no.~7, (Jul, 2014) }.
  \url{https://doi.org/10.1007%2Fjhep07%282014%29079}.

\bibitem{triplet_dm}
A.~Belyaev, G.~Cacciapaglia, J.~Mckay, D.~Marin, and A.~R. Zerwekh, ``{Minimal
  Spin-one Isotriplet Dark Matter},''
  \href{http://dx.doi.org/10.1103/PhysRevD.99.115003}{{\em Phys. Rev. D}
  {\bfseries 99} no.~11, (2019) 115003},
  \href{http://arxiv.org/abs/1808.10464}{{\ttfamily arXiv:1808.10464
  [hep-ph]}}.

\bibitem{CMS-PAS-EXO-14-012}
{\bfseries CMS} Collaboration, ``{Search for long-lived particles that decay
  into final states containing two muons, reconstructed using only the CMS muon
  chambers},'' tech. rep., CERN, Geneva, 2015.
\newblock \url{http://cds.cern.ch/record/2005761}.

\bibitem{susylims_dy}
{\bfseries ATLAS} Collaboration, G.~Aad {\em et~al.}, ``{Search for electroweak
  production of charginos and sleptons decaying into final states with two
  leptons and missing transverse momentum in $\sqrt{s}=13$ TeV $pp$ collisions
  using the ATLAS detector},''
  \href{http://dx.doi.org/10.1140/epjc/s10052-019-7594-6}{{\em Eur. Phys. J. C}
  {\bfseries 80} no.~2, (2020) 123},
  \href{http://arxiv.org/abs/1908.08215}{{\ttfamily arXiv:1908.08215
  [hep-ex]}}.

\end{thebibliography}\endgroup
\addcontentsline{toc}{chapter}{\bibname}
\end{document}